\documentclass{aa}  

\usepackage{graphicx}
\usepackage{txfonts}
\usepackage{hyperref}

\begin{document}

   \title{The vanishing of the long photometric cycle in AU Monocerotis}


   \author{L. Celed\'on
          \inst{1}
          \and
          R.E. Mennickent\inst{2}
          \and
          D. Barr\'ia\inst{3}
          \and
          J. Garc\'es\inst{2}
          \and
          M. Jurkovi\'c\inst{4}
          }

   \institute{Instituto de F\'isica y Astronom\'ia, Universidad de Valpara\'iso, Av. Gran Breta\~{n}a 1111, Valpara\'iso, Chile.\\
              \email{lientur.celedon@postgrado.uv.cl}
              \and
              Departamento de Astronom\'ia, Universidad de Concepci\'on, Av. Esteban Iturra s/n Casilla 160-C, Concepci\'on, Chile.
              \and
              Centro de Investigaci\'on en Ciencias del Espacio y F\'isica Te\'orica, Universidad Central de Chile, Av. Francisco de Aguirre 0405, La Serena, Chile.
              \and
              Astronomical Observatory, Volgina 7 11000, Belgrade, Serbia.
             }

   \date{Received month day, year; accepted month day, year}
 
  \abstract
   {
   Double periodic variables (DPVs) are a group of semi-detached interacting binaries that exhibit a long photometric cycle with an average length of approximately 33 times the orbital period of the system.
   It has been proposed that this long photometric cycle originates from a modulated mass transfer rate from the donor star, which itself is driven by an internal magnetic dynamo within the donor. One of the most well-studied DPVs in the Milky Way is AU Monocerotis (AU Mon).
   }
   {
   We aim to enhance our understanding of the origin of the long photometric cycle in AU Mon by characterising its behaviour through the analysis of available photometric data from several databases and surveys.
   }
   {
   We summarise previous findings on the system and analyse its published multi-wavelength photometry from different sources, covering 46.3 years, to study the variability of its light curve.
   }
   {
   We find that the orbital period has remained constant over recent decades, but the long cycle of approximately 417 days vanished around 2010. 
   From an O-C analysis, we conclude that the system is experiencing a change in its orbital period of no greater than $0.038\pm0.040$ s\,yr$^{-1}$, and thus, imposing a value of $2\times10^{-8}$ M$_\odot$\,yr$^{-1}$ for $\dot{M}$ in a fully conservative mass transfer regime. 
   The disappearance of the long cycle is more evident in the V filter than in the Ic filter. In the latter, a small amplitude variation related to the long cycle is still detected.
   A time-series analysis of the disentangling light curve in the Ic filter shows a transient periodicity of approximately $1\,910$ days lasting at least $2\,000$ days before it also disappears around the year 2020.
   An analysis of the available AAVSO Photometric All-Sky Survey light curves around the year 2013 shows a strong periodicity at approximately 280 days, which appears to be stronger in the Z filter.
   }
   {
   We report what is the second observation of the sudden disappearance of the long cycle in a DPV, after the Galactic DPV TYC 5353-1137-1. The disappearance of the long cycle in AU Mon is a strong constraint for current models that aim to explain the long cycle in DPVs.
   }

   \keywords{stars: binaries: close --
             individual object: AU Mon --
             techniques: photometric
            }

   \maketitle

\section{Introduction}

Interacting binaries are systems in which one of the stars (the donor star) transfers matter to its companion (the gainer), thereby affecting the evolution of both stars in the process \citep{Pringle_1985}.
As a consequence of mass transfer, several physical processes can take place in the system, such as accretion disc formation, mass accretion, and the ejection of matter from the system.
This makes interacting binaries excellent laboratories for studying complex astrophysical phenomena related to mass exchange between stars, systemic mass loss from the system, the physics of accretion discs, and binary star evolution.

There are several subclasses of interacting binaries. These have mainly been classified based on the evolutionary stage of their components (main sequence or evolved stars), and on the geometrical and physical conditions of the binary system, with important parameters including their orbital periods, and the mass ratio, $q=M_{2}/M_{1}$, between components.
One of these subclasses, first reported by \cite{Mennickent+2003} in the Magellanic Clouds, are the so-called double periodic variables (DPVs).
These interacting binaries are composed of a giant, cold donor star of spectral type A, F, or K, which has filled its Roche lobe and transfers matter to its more massive and hotter B-type companion via Roche lobe overflow in case B of mass transfer \citep[e.g.][]{Rosales+2023A&A...670A..94R}.
The distinctive characteristic of DPVs is the existence of an additional long-term brightness variation, whose period ($P_\mathrm{l}$) is correlated with the orbital period ($P_\mathrm{o}$) of the system by the relation $P_\mathrm{l}\sim33\,P_\mathrm{o}$ \citep{Mennickent+2003, Mennickent2017, Mennickent2022}.
However, recent data published in the fourth phase of the Optical Gravitational Lensing Experiment (OGLE) project suggest a bimodal distribution of the $P_\mathrm{l}/P_\mathrm{o}$ relation, with a secondary peak around 18 \citep{Glowacki+2025arXiv250315596G}.
Such bimodality may indicate different evolutionary paths for some DPVs or hidden physical processes that are not yet understood \citep[see for example, the well-known `period gap' in Cataclysmic Variables from which the theory of magnetic braking was developed; ][]{Spruit&Ritter1983A&A...124..267S}.
To date, approximately 300 DPVs have been discovered in the Magellanic Clouds and the Milky Way. Their orbital period distribution shows a peak around nine days, with some systems showing orbital periods of approximately 1 and 100 days \citep{Mennickent2017, Mennickent2022, Rosales+2023A&A...670A..94R}.
Long-term variability, on the other hand, typically exhibits an amplitude of $\Delta \mathrm{I}\sim0.2$\,mag and periods between $150$ and $1\,000$ days \citep{Mennickent2005}.

The nature of the physical process behind the long photometric cycle remains under debate. The most recently proposed scenario suggests that it is the result of a modulated mass transfer rate ($\dot{M}$) due to a magnetic dynamo within the secondary star \citep{Schleicher2017}. Other interpretations, based on deep studies of individual systems, have been proposed in the past. These include cyclic pulsations of the donor star in the DPV AU Monocerotis (AU Mon), which resulted in a change in the $\dot{M}$ value \citep{Peters1994}; cycles of mass loss from the system, probably feeding a circumbinary disc at the DPV OGLE5155332-6925581 \citep{Mennickent2008}; and disc winds at the DPV V393 Scorpii \citep{Mennickent2012b}, among others. 
More recently, following a multi-wavelength spectroscopic study of AU Mon using optical and UV archival data, \cite{Armeni_2022} suggest that the long-term variability is driven by the accretion disc, which alters its structure during the `faint' and `bright' states of the long-term cycle. Despite these previous efforts, no conclusive interpretation of the long cycle has been provided so far, and further studies are needed to clarify this phenomenon.

Interestingly, long-term brightness variations do not appear to be strictly periodic, as pointed out by \cite{Mennickent2004} and later supported by the discovery of a system that shows a shortening of its long-term period by $\sim20\%$ over several cycles \citep{Mennickent2005, Mennickent+2019MNRAS.487.4169M}. 
The long period was reported to decrease continuously from 350 to 218 days over 13 years in the system OGLE-LMC-DPV-065 \citep{Mennickent+2019MNRAS.487.4169M}, while OGLE-BLG-ECL-157529 shows that its long cycle decreases in amplitude and length along 18.5 years \citep{2020A&A...641A..91M}.
Furthermore, in the galactic DPV TYC 5353-1137-1, the long period was reported to disappear for approximately $1\,500$ days before reappearing as before \citep{Rosales&Mennickent2018IBVS.6248....1R}.
It is worth noting that previous studies took advantage of the photometric data density provided by the OGLE programme in the Magellanic Clouds and the Bulge of the Galaxy. There are no similar wide surveys in the Galaxy besides the Bulge, and long-term, high-cadence data on Galactic DPVs are fragmentary and not robust.
Remarkably, despite substantial evidence for cyclic mass loss having been reported in some well-known DPV systems, no significant change in their orbital periods has been detected \citep{Mennickent2012a, Mennickent2014, Garrido2013, Barria2013}. This contrasts with the theoretical framework, in which a change of the orbital period for high $\dot{M}$ values is expected in a conservative mass transfer scenario.

Among the known galactic DPVs, AU Mon is one of the most well-studied systems. It is an eclipsing, bright ($\mathrm{V}=8.4$\,mag), double-line spectroscopic binary with an orbital period of 11.1130374(1)\,d \citep[][D10 from here on]{Desmet2010}. The physical parameters of the system have been determined independently by light-curve modelling \citep{Desmet2010, Djurasevic2010} and through spectroscopic analysis \citep{Sahade+1997, Vivekananda&Sarma1998}. The derived system parameters, however, differ slightly between studies.
Analysis of the AU Mon light curve reveals the presence of a long photometric modulation with a period of $\sim$417\,d ($P_\mathrm{l}/P_\mathrm{o} \sim 37.5$) and an amplitude of 0.2\,mag in the V band \citep{Lorenzi80, Lorenzi1985, Desmet2010}.

To date, and to our knowledge, possible long-term variations in the long cycle of AU Mon have not been investigated. 
In this work, we present a reanalysis of the available archival photometry of AU Mon spanning over 40 years, collected from different ground-based instruments and telescopes, to investigate the stability of the orbital and long cycle of the system, both in terms of photometric amplitude and cycle-length duration.

\section{AU Monocerotis} \label{sec:AU_Mon}

\begin{table*}
    \centering
    \caption{Compilation of physical parameters of the AU Mon system from the literature.}
    \label{tab:aumon_recompilation}
    \resizebox{0.95\textwidth}{!}{
    \begin{tabular}{lrrrrr}
    \hline \hline \noalign{\smallskip}
    & Sahade et al., & Vivekananda Rao\&Sarma, & Desmet et al., & Djurasevic et al., & Atwood-Stone et al., \\
    & 1997 & 1998 &  2010 &  2010 &  2012 \\
    \hline \noalign{\smallskip}
     $M_\mathrm{h}$ [M$_\odot$] & 6.1$\pm$0.61 & 5.93$\pm$0.31 & 6.37$^{+2.18}_{-1.12}$ & 7.0$\pm$0.3 & 7.0\tablefootmark{a} \\
     $M_\mathrm{c}$ [M$_\odot$] & 1.8$\pm$0.53 & 1.18$\pm$0.16 & 1.17$\pm$0.19 & 1.2$\pm$0.2 & 1.2\tablefootmark{a} \\
     $R_\mathrm{h}$ [R$_\odot$] & -- & 5.28$\pm$0.16 & 7.15$^{+5.77}_{-2.92}$ & 5.1$\pm$0.5 & 5.1\tablefootmark{a} \\
     $R_\mathrm{c}$ [R$_\odot$] & -- & 10.04$\pm$0.74 & 9.7$\pm$0.6 & 10.1$\pm$0.5 & 10.1\tablefootmark{a} \\
     $T_\mathrm{h}$ [K] & -- & 15\,500$\pm$100\tablefootmark{a} & 15\,000$\pm$2\,000 & 15\,890$\pm$400 & 17\,000 \\
     $T_\mathrm{c}$ [K] & -- & 6000$\pm$40 & 5\,750$\pm$250 & 5\,750 & 5\,750\tablefootmark{a} \\
     $q$ & 0.3$\pm$0.09 & 0.1985\tablefootmark{a} & 0.17$\pm$0.03 & 0.17$\pm$0.03\tablefootmark{a} & 0.17\tablefootmark{a} \\
     $P_\mathrm{o}$ [d] & 11.1130371\tablefootmark{a} & 11.1130371\tablefootmark{a} & 11.1130374 & 11.1130374\tablefootmark{a} & 11.1130374\tablefootmark{a} \\
     $e$ & 0.06$\pm$0.02 & 0.0\tablefootmark{a}& 0.01$\pm$0.03 & 0.0\tablefootmark{a} & 0.0\tablefootmark{a} \\
     $i$ [deg] & -- & $78.74\pm0.06^{\circ}$ & $78.6\pm0.6^{\circ}$ & $80.1\pm0.6^{\circ}$ & $80.1^{\circ}$\tablefootmark{a} \\
     $r_\mathrm{disc}$ [R$_\odot$] & -- & -- & -- & 12.7$\pm$0.6 & 23$\pm$2 \\
     $T_\mathrm{disc}$ [K] & -- & -- & -- & 15\,870\tablefootmark{b} & 14\,000$\pm$2\,000\tablefootmark{b} \\
    \hline
    \end{tabular}
    }
    \tablefoottext{a}{Assumed value}
    \tablefoottext{b}{Refers to the disc inner radius}
    \tablefoot{
    Masses, radius, and temperature of the hot and cool components, the mass ratio of the system, its period, eccentricity and inclination, together with estimations of the accretion disc size and temperature, are provided.
    }
\end{table*}

AU Mon is an eclipsing, Algol-type interacting binary composed of a hot B-type primary and a cooler F- or G-type secondary star.
Its periodic nature was first identified by \citep{Hoffmeister1931}, while the first ephemerids were provided by \citet{Floria1937}, who derived an orbital period of 11.11 days.
The binary nature of AU Mon was confirmed following the first spectroscopic studies of the system, carried out by \citet{Sahade&Cesco1945}.
From their analysis, they determined the semi-amplitude of the radial velocity curve and classified the components as B5 (hot) and F0 (cold).
Subsequent studies refined the orbital period by analysing the system's light curve \citep[see, for example][]{Lause1949, Lorenzi80}.
The physical parameters of the system were revised using spectroscopic studies, which revealed a double-peak emission profile around H$\alpha$, indicative of the presence of an accretion disc \citep[e.g. ][]{Popper1962PASP...74..129P, Sahade&Ferrer1982PASP...94..113S, Sahade+1997, Vivekananda&Sarma1998}.

Significant progress in clarifying the system's physical parameters was made after photometry from the Convection, Rotation and planetary Transits (CoRoT) space mission was published. This dataset allowed D10 to obtain the most precise measurement of the orbital period of the system, corresponding to 11.1130374(1)\,d based on the time-series analysis of four consecutive eclipses.
By combining the CoRoT photometric data with additional spectroscopic data, the authors determined the orbital and physical parameters of the system, arguing for a spectral type G instead of F for the cooler star.
The modelling attempted by D10 did not consider the presence of an accretion disc in the system. This omission motivated \citet{Djurasevic2010} to model the CoRoT light curve, this time including an optically thick accretion disc around the hot star. Their best model considered three active regions on the disc edge: one corresponding to the hot spot resulting from the interaction between the disc and the stream, and the other two to bright spots.
The disc parameters were later modelled by \citet{Atwood2012} based on the system's H$\alpha$ profile, corroborating the presence of active regions within the disc. They found that the H$\alpha$ profile could be explained by the inclusion of an accretion disc, a stream of matter from the donor star, and a hot spot originating in the impact zone between the disc and the stream.
The results of both models are in overall good agreement, with the main difference lying in the extension of the accretion disc, where a discrepancy factor of two is observed. This difference is attributed to the different approaches, with the work of \citet{Djurasevic2010} being more sensitive to the inner parts of the disc, while \citet{Atwood2012} traced the outermost parts.

The position of AU Mon in the radius-mass ratio (r-q) diagram shows that the system can host a large accretion disc around the hot star, fed by Roche-lobe overflow from the cool star \citep{Richards&Albright1999ApJS..123..537R, Mennickent+2016MNRAS.455.1728M}. Thus, the disc radius found by \citet{Atwood2012} is a feasible possibility.
A summary of the different parameters derived from these studies is presented in Table~\ref{tab:aumon_recompilation}.
In addition to the accretion disc, the position of the system in the r-q diagram also predicts that AU Mon is a tangential impact system. 
In this kind of system, the falling stream of matter from the donor star is very efficient at transferring angular momentum to the gainer star, effectively increasing its rotation.
Several observations have confirmed that the hot star in AU Mon shows an asynchronous ratio (observed projected rotation velocity over synchronous rotation) of approximately five, while the secondary rotates synchronously \citep{Glazunova+2008AJ....136.1736G, Armeni_2022}.

The system's long cycle was first identified by \citet{Cerruti-Sola&Lorenzi1977}, who observed that the entire light curve of the system shifted up and down in brightness.
Later, \citet{Lorenzi80} determined that these variations occur cyclically with a period of $\sim$411\,d and an amplitude of $\sim$0.2\,mag in the V band.
The origin of this long-cycle variation remains unclear.
Based on the variability observed in the far-ultraviolet (FUV) spectra, \citet{Peters1994} proposed that the long cycle is regulated by a variable mass transfer rate, $\dot{M}$, coming from the cool star, which feeds circumstellar material around the hot star, effectively creating an extra source of continuum emission.
This idea was revisited by \citet{Armeni_2022}, who instead argued in favour of a change in the height of the accretion disc, as a result of the variable mass transfer rate, as a possible explanation for the long-cycle variation.

The physical reason behind the variable $\dot{M}$ could be associated with pulsations from the cool star. This idea was tested by searching for possible pulsations in the CoRoT photometry, finding conclusive evidence that supports the hypothesis of pulsations in the B-type hot star, although their origin is undefined, as well as solar-type pulsations in the cool star \citep{Desmet2010, Djurasevic2010}. 
The optical spectra of the system, particularly the H$\rm\alpha$ line, also support the presence of circumstellar material that varies with the long cycle, as the central absorption of the line becomes deeper and broader during the low states of the long cycle \citep{Barria&Mennickent2011, Atwood2012}.
Conversely, \citet{Djurasevic2010} attempted to reproduce the long-cycle variation by varying the physical parameters of the accretion disc, but without success, leading them to conclude that the most likely origin for the long cycle lies in a circumbinary disc around the system.

\citet{Mennickent2014} investigated the evolutionary stage of AU Mon and concluded that the system is likely $\sim$196\,Myrs old and is currently undergoing case-B mass transfer.
Their results also indicate that the system experiences a relatively high $\dot{M}$ of 7.6$\times$10$^{-6}$\,M$_\odot$\,yr$^{-1}$.
This value is incompatible with the constancy of the reported orbital period under the assumption of a conservative mass-transfer rate, leading the authors to conclude that a significant part of the transferred angular momentum is lost to the interstellar medium via outflows.
This interpretation is supported by the observation of P-Cygni profiles and discrete absorption components in Si and Al lines in the FUV spectrum \citep{Armeni_2022}.

\section{Data acquisition} \label{sec:data_acquisition}

AU Mon photometric data have been collected by several sources over the past several decades, which, when combined, provide extensive temporal coverage for studying the system's variability.
For this study, we considered photometric data from several sources.
A summary of the collected data, chronologically sorted by the start of their observations, is presented in Table~\ref{tab:summary_photometry}. The table includes the different instruments and filters employed, the effective wavelength of the filter used ($\lambda_\mathrm{eff}$), the start and end of each observation, the total number of collected data points ($N$), and the photometric mean error ($\overline{\mathrm{m}_\mathrm{err}}$) for each dataset.
The epochs listed are in Modified Julian Date (MJD) with heliocentric correction (that is, $\mathrm{HJD}-2\,400\,000.5$).

The listed data include the 2\,616 V-band measurements between 1976 and 1979 provided by \citet{Lorenzi80}. Additional V band data were contributed by the American Association of Variable Star Observers (AAVSO) \footnote{\href{https://www.aavso.org/}{AAVSO web page}}, including 28 points from 1976–2003; Northern Sky Variability Survey (NSVS) \citep{NSVS2004} contributing with 50 points from 1999–2000; the third release of All Sky Automated Survey (ASAS-3) \footnote{\href{https://www.astrouw.edu.pl/asas/}{ASAS web page}} \citep{Pojmanski2002} with 336 points from 2001–2009; All-Sky Automated Survey for Supernovae (ASAS-SN) \footnote{\href{https://asas-sn.osu.edu/}{ASAS-SN web page}} with 266 points from 2014–2018; and Kamogata/Kiso/Kyoto Wide-field Survey (KWS) \citep{Maehara2015_KWS} \footnote{\href{https://jaxa.repo.nii.ac.jp/records/1963}{Available in JAXA repository.}} providing 1\,155 measurements.
The Bochum Galactic Disc Survey \footnote{\href{https://galacticdiscsurvey.space/index.html}{Bochum survey web page}} \citep{Haas+2012, Haas+2015} provided 96 i-band observations between 2010 and 2014, while KWS contributes a further 954 observations in the Ic band.
The ESA’s Hipparcos mission \citep{Hipparcos97, Hipparcos2007} contributed with an additional 90 observations using a wide visual filter, while APASS \footnote{\href{https://www.aavso.org/apass}{APASS web page}} \citep{Smith2010, Henden2011} contributed several hundred multi-band measurements (2011–2016).
Although KWS provides up-to-date photometry for AU Mon, in this work we considered data only up to the end of 2023 \footnote{\href{http://kws.cetus-net.org/~maehara/VSdata.py}{Access to the data}}.

\begin{table}[]
    \centering
    \caption{Summary of the photometric datasets used in this work.}
    \label{tab:summary_photometry}
    \resizebox{1.0\columnwidth}{!}{
    \begin{tabular}{lcclll}
        \hline\hline\noalign{\smallskip}
        Source & Filter & $\lambda_\mathrm{eff}$ & MJD range & $N$ & $\overline{\mathrm{m}_\mathrm{err}}$\\
               &        & [nm]                   &           &     &  \\
        \hline\noalign{\smallskip}
        Lorenzi1980 & V  & 540 & 42790 - 43880 & 2613 & --    \\
        AAVSO       & V  & 540 & 43079 - 52721 & 28   & --    \\
        Hipparcos   & Hp & 503 & 47987 - 49056 & 90   & 0.013 \\
        NSVS        & V  & 540 & 51504 - 51630 & 50   & 0.010 \\
        ASAS-3      & V  & 540 & 52033 - 55165 & 336  & 0.040 \\
        Bochun      & i  & 749 & 55467 - 56984 & 95   & 0.007 \\
        KWS         & V  & 540 & 55525 - 59649 & 1155 & 0.015 \\
        APASS       & B  & 442 & 55827 - 57421 & 678  & 0.002 \\
                    & V  & 540 & 55827 - 57421 & 678  & 0.002 \\
                    & g  & 470 & 55836 - 57421 & 415  & 0.014 \\
                    & r  & 618 & 55836 - 57421 & 408  & 0.003 \\
                    & i  & 749 & 55836 - 57421 & 411  & 0.004 \\
                    & z  & 899 & 55949 - 56254 & 19   & 0.004 \\
                    & Z  & 876 & 56199 - 57421 & 430  & 0.013 \\
        KWS         & Ic & 800 & 56578 - 59649 & 954  & 0.025 \\
        ASAS-SN     & V  & 540 & 57009 - 58378 & 266  & 0.020 \\
        \hline
    \end{tabular}
    }
\end{table}

\section{Photometric analysis}

\begin{figure*}
    \centering
    \includegraphics[width=1.0\textwidth]{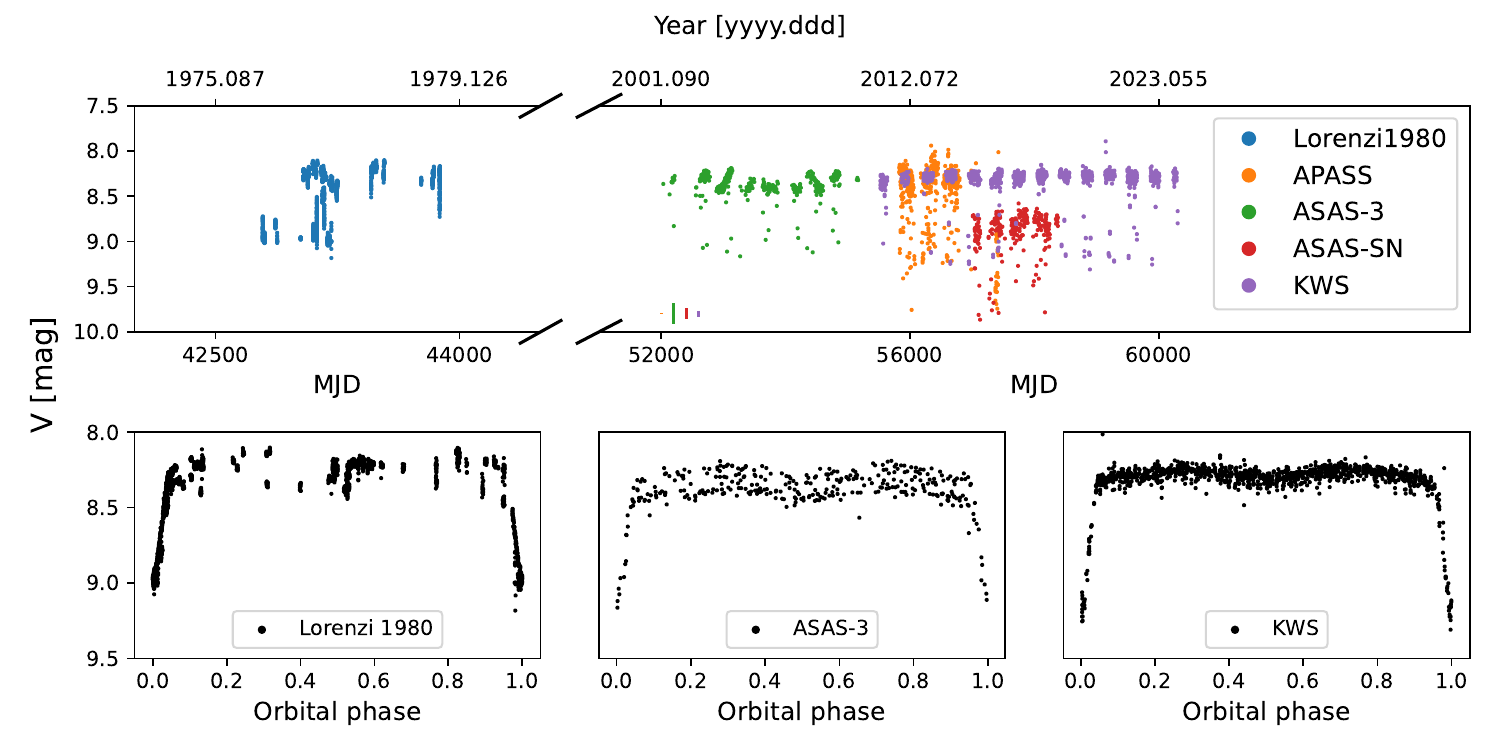}
    \caption{
    V band photometry of AU Mon from Lorenzi1980, ASAS-3, ASAS-SN, APASS, and KWS.
    Upper panel: V magnitude versus MJD (bottom axis) and year (top axis, in ordinal format).
    The mean photometric error for each dataset is indicated by error bars at the bottom of the upper panel.
    Bottom panels: Lorenzi1980, ASAS-3, and KWS photometry phased with the orbital period of 11.1130374 d. 
    Sinusoidal modulation associated with the long-cycle period is not evident in the KWS photometry, while the orbital period appears to have remained unaltered over the past decades.
    }
    \label{fig:photometry1}
\end{figure*}

Our analysis begins with an examination of the available photometry from the different dataset collections.
The photometry in the V band is presented in the upper panel of Fig.~\ref{fig:photometry1}.
The photometry clearly shows the modulation associated with the long cycle in the Lorenzi1980 dataset and ASAS-3 datasets, but apparently disappears in the more recent KWS data. 
We excluded the AAVSO and NSVS datasets from Fig.~\ref{fig:photometry1} due to their poor photometric quality and low number of data points.
The light curve for each dataset is presented in Appendix~\ref{sec:appendix_lcs}, including its phase curve against the orbital and long period reported in D10. From this, it is clear that the quality of AAVSO, NSVS, and APASS-z is insufficient for a reliable analysis.

However, these same datasets reveal the constancy of the orbital period. This phenomenon is illustrated in the bottom panels of Fig.~\ref{fig:photometry1}, where we present the data corresponding to the Lorenzi1980, ASAS-3, and KWS datasets phased with the orbital period and ephemeris reported by D10. In all cases, the orbital light curve is consistent with the reported values, revealing that this has remained stable over the past several decades.

\subsection{Orbital cycle} \label{sec:orbital_cycle}

To assess the stability of the orbital period over time, we performed a time-series analysis of each available photometric dataset. 
We employed two different statistical tools to identify dominant periods (or frequencies) in the data: the generalized Lomb-Scargle (GLS) periodogram, implemented in the \texttt{PyAstronomy.pyTiming}\footnote{\href{https://pyastronomy.readthedocs.io/en/latest/pyTimingDoc/pyPeriodDoc/gls.html}{PyAstronomy.pyTiming}} \texttt{Python} package \citep{Zechmeister2009} and the phase dispersion minimization (PDM) method implemented in the \texttt{pdmpy}\footnote{\href{https://py-pdm.readthedocs.io/en/latest/}{pdmpy}} \texttt{Python} package \citep{Stellingwerf1978}.
Both methods are suitable for detecting periodic signals in a dataset, and have been widely used in the analysis of DPVs \citep[e.g. ][]{Mennickent+2020A&A...641A..91M, Rosales+2023A&A...670A..94R}.
The main difference between the methods is that GLS assumes a periodic signal with a sinusoidal shape, whereas PDM does not make any assumption about the shape of the periodic signal. By comparing the results of both methods, we evaluated the significance of the detected signal.

Our results are presented in Table~\ref{tab:orbital_period_measurements}, where we list the period corresponding to the strongest signal found by each method in a window between 9 and 13 days.
In the case of the GLS method, the uncertainties associated with the determination of the sinusoidal period were obtained from the analysis of the goodness of fit.
In contrast, the PDM method does not provide an estimate of the uncertainty associated with the detected frequencies. In this case, the uncertainties were quantified using the standard deviation of the Gaussian fit applied to the most prominent peaks in the frequency spectra.
Additionally, the PDM method implemented requires that the input dataset contains at least 100 data points. Thus, datasets containing fewer than 100 observations were not considered for the PDM analysis (no reported PDM period in Table~\ref{tab:orbital_period_measurements}).
The results indicate that both methods are consistent with each other within the uncertainties and that the periodic signal is also consistent with the orbital period provided by D10, supporting the stability of the orbital period across the years.
There are, however, some datasets that show inconsistencies compared to the rest, namely APASS-z ($P_\mathrm{o} \sim 10.1 \,\mathrm{d}$), Hipparcos ($P_\mathrm{o} \sim 11.8 \,\mathrm{d}$), and NSVS ($P_\mathrm{o} \sim 11.8 \,\mathrm{d}$).
In the case of APASS-z and NSVS, we attribute the discrepancy to the poor photometric quality of the dataset (Figs.~\ref{fig:appendix_apassz} and ~\ref{fig:appendix_nsvsV}), while in the case of the Hipparcos dataset, the difference can be attributed to a combination of poor sampling and contamination from the long cycle.
We noted that after the long cycle is disentangled from the light curve (Sect.~\ref{sec:long_cycle}), a prominent peak corresponding to the orbital period of $\sim$11.113 days appeared with the same power as the 11.77 days signal.

\begin{table}[]
    \centering
    \caption{Orbital period measurements for the different datasets used.}
    \label{tab:orbital_period_measurements}
    \resizebox{1.0\columnwidth}{!}{
    \begin{tabular}{lcrrr}
        \hline\hline \noalign{\smallskip}
        Observer & Filter & $P_\mathrm{GLS}$ & $P_\mathrm{PDM}$ & $\overline{P}$ \\
        \hline \noalign{\smallskip}
        Lorenzi1980 & V  & 11.084(003) & 11.100(050) & 11.090(030) \\
        AAVSO       & V  & 11.087(001) & --          & 11.087(001) \\
        Hipparcos   & Hp & 11.770(010) & --          & 11.770(010) \\
        NSVS        & V  & 11.800(200) & --          & 11.800(200) \\
        ASAS-3      & V  & 11.110(004) & 11.110(010) & 11.110(005) \\
        Bochun      & i  & 11.110(020) & --          & 11.110(020) \\
        KWS         & V  & 11.114(002) & 11.112(009) & 11.113(005) \\
        APASS       & B  & 11.114(009) & 11.110(040) & 11.110(020) \\
                    & V  & 11.103(009) & 11.110(040) & 11.107(020) \\
                    & g  & 11.100(020) & 11.120(050) & 11.110(030) \\
                    & r  & 11.090(010) & 11.110(050) & 11.100(030) \\
                    & i  & 11.100(010) & 11.130(050) & 11.115(025) \\
                    & z  & 10.080(090) & --          & 10.080(090) \\
                    & Z  & 11.120(030) & 11.120(050) & 11.120(030) \\
        KWS         & Ic & 11.114(004) & 11.112(009) & 11.113(005) \\
        ASAS-SN     & V  & 11.110(010) & 11.110(030) & 11.110(020) \\
        \hline
    \end{tabular}
    }
    \tablefoot{
    The values obtained by each method and their mean value are provided with their respective uncertainties, expressed as the last significant digit in parentheses.
    In the case that the dataset contains fewer than 100 points, the PDM method could not be implemented (see main text for details).
    }
\end{table}

\subsection{Observed minus calculated analysis}

An additional technique to determine whether the orbital period of an eclipsing binary has changed over time is the so-called observed minus calculated diagram (O-C) \citep{Kalimeris94}. This diagram shows the differences between the observed times of the primary minima in a light curve and the predicted times, assuming a constant period. If the assumed period corresponds to the true one and has remained constant, the data should be distributed around zero in the O-C diagram. However, if the orbital period experiences a continuous change with time, as expected in a binary system with conservative mass transfer, a parabolic shape should appear in the diagram. The quadratic coefficient of the polynomial fit provides the orbital period change rate, $\dot{P_\mathrm{o}}$, experienced by the binary, according to

\begin{equation}
    \dot{P_{\mathrm{o}}} = 2\,C_\mathrm{q}/P_\mathrm{o},
\end{equation}

where $C_\mathrm{q}$ corresponds to the quadratic polynomial coefficient derived from the O-C fit.
To construct an O-C diagram for AU Mon, we used the period and ephemeris reported by D10, as well as the O-C values provided by the authors of CoRoT and previous observations.

To maximise the quality of the O-C analysis, it is important to identify the times of minima with high precision. This is usually achieved by analysing light curves with high time cadence, such as the CoRoT light curve analysed by D10.
The datasets analysed in this work do not have such a high cadence; thus, the identification of minima becomes less straightforward.
A selection criterion was applied based on the orbital phase (with respect to the D10's orbital period) and magnitude to identify data points belonging to the primary eclipse, that is, the minimum.
The search for primary eclipse points was restricted to a region between 9.0 and 9.4 mag, and an orbital phase between 0.97 and 0.03.
The general stability of the orbital period was demonstrated in Sect.~\ref{sec:orbital_cycle}, although the uncertainties allow for possible variations on the order of minutes.
The criteria implemented for selecting primary eclipse points guarantee the selection of points belonging to the primary eclipse, even if a small variation in the orbital period is present.

To maintain consistency, only datasets using the V filter were considered for the O-C analysis.
For each point identified as belonging to a primary eclipse in each dataset, its cycle number--that is, the number of orbital cycles that have passed between the ephemeris and the observed time--was determined.
This allowed for the calculation of the expected time of the primary eclipse and the subsequent construction of the O-C diagram.
Due to the high dispersion observed in the photometric data and its low cadence, it is difficult to determine the time of each eclipse with precision.
Therefore, standard techniques such as the use of polynomials to determine the time of minima and their associated uncertainty are not suitable.
Consequently, an arbitrary value of 0.1 days was assigned as the uncertainty in the time of each minimum.
This value was chosen because it corresponds approximately to half the duration of the primary eclipse.
This is a conservative value for the uncertainty, assuming that the actual primary eclipse ocurred within this window.
For the O-C values provided by D10 before the CoRoT observations (prevCoRoT), an arbitrary uncertainty value of 0.2 days was adopted to reduce their weighting relative to the more recent data.

\begin{figure}
    \centering
    \includegraphics[width=1.0\columnwidth]{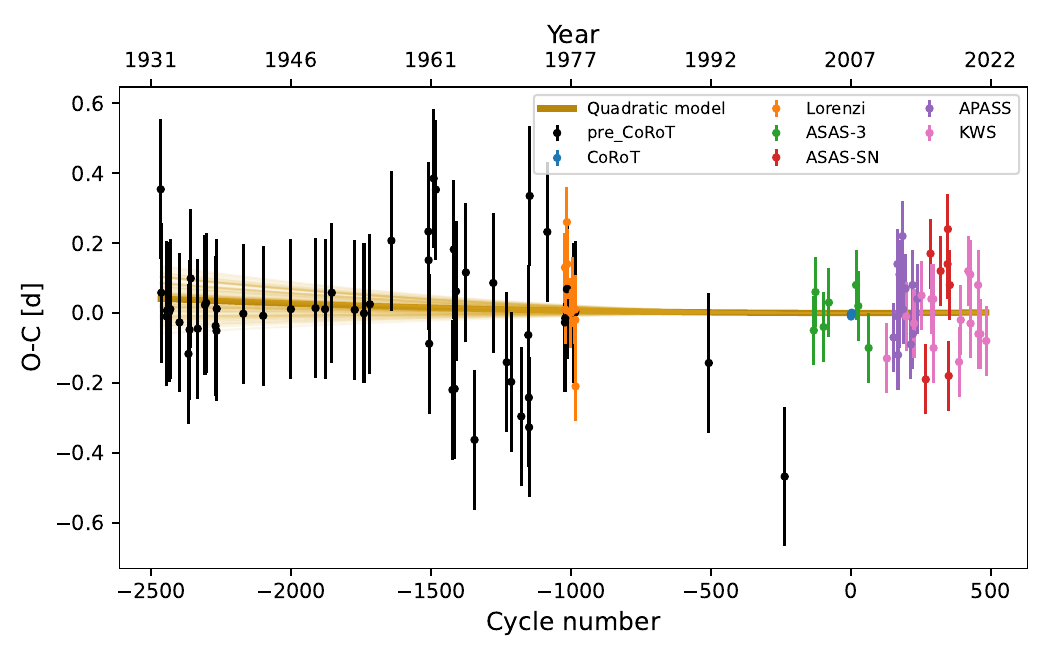}
    \caption{
    Observed-calculated (O-C) diagram for values published by D10, including those derived from CoRoT data and previous observations (prevCoRoT), along with the values derived in this work. 
    The best quadratic fit to the data is shown in dark gold, with lighter lines representing 100 random models used to illustrate the uncertainty in the fit parameters.}
    \label{fig:O-C_diagram}
\end{figure}

The resulting O-C diagram is shown in Fig.~\ref{fig:O-C_diagram}. An initial inspection of the diagram shows values scattered around zero, with deviations no larger than 0.4 days, supporting the idea of a constant orbital period over the years.
However, a small variation in the orbital period is still plausible. We therefore fitted the O-C data using three different models: a pure linear fit, a pure quadratic fit, and a combination of linear and quadratic models. Within the O-C context, a linear fit represents a constant orbital period, but with a true value different from the one used to construct it (the D10 orbital period in this case). A pure quadratic fit indicates a constant variation in the orbital period, while a combination of models indicates a mix between these two effects.

The coefficients for the best fit at each model were determined using the non-linear least squares minimisation method implemented in \texttt{SciPy}, with the respective uncertainties in each coefficient determined from the resulting covariance matrix.
The best coefficient values are provided in Table~\ref{tab:stan_results} together with their corresponding $\dot{P_\mathrm{o}}$ values.
Based on the standard error of the estimate (SEE) values, the quadratic model provides the best fit, albeit by only a small margin, suggesting a constant increase of $\sim$4$\times$10$^{-2}$ s yr$^{-1}$ in the orbital period of AU Mon over the last century. However, it should be noted that the uncertainty in the derived coefficients is sufficiently large to allow a constant orbital period, or even a decrease.

\begin{table}[]
    \centering
    \caption{Results from the O-C analysis.}
    \label{tab:stan_results}
    \resizebox{1.0\columnwidth}{!}{
    \begin{tabular}{c c c c c}
        \hline\hline \noalign{\smallskip}
        Model & $C_\mathrm{l}$ & $C_\mathrm{q}$ & $\dot{P_{\mathrm{o}}}$ & SEE \\ 
        & [$10^{-5}$ d] & [$10^{-9}$ d] & [$10^{-2}$ s\,yr$^{-1}$]& \\ 
        \hline \noalign{\smallskip}
        Linear             & -1.0$\pm$1.3 & ---         & ---           & 0.1428 \\
        Quadratic          & ---         & 6.7$\pm$7.1 & 3.8$\pm$4.0 & 0.1427 \\ 
        Linear + Quadratic & 0.5$\pm$3.2 & 9.3$\pm$17.0 & 5.3$\pm$9.7 & 0.1433 \\
        \hline
    \end{tabular}
    }
    \tablefoot{
    The linear ($C_\mathrm{l}$) and quadratic ($C_\mathrm{q}$) coefficients with their respective uncertainties, their corresponding orbital period variation $\dot{P_{\mathrm{o}}}$, and the standard error of estimates (SEE) between the observed and model data are presented.
    }
\end{table}

\subsection{Long cycle} \label{sec:long_cycle}

To investigate the stability of the long cycle, we disentangled the orbital light curve from the raw photometry, allowing us to better analyse the long-term variability. Hereafter, we refer to the light curve resulting from removing the orbital periodicity as the disentangled light curve. The disentangling was performed by adjusting a fundamental frequency and a series of $n$ harmonics to the raw data using a Fourier series, before subtracting this signal from the original data. This method has previously been implemented to successfully disentangle the long cycle from the orbital light curve in DPVs \citep[e.g.][]{Mennickent2008, Mennickent2012a, Garces2019}. The fundamental frequency corresponds to that of the orbital period (0.0899844 d$^{-1}$), and a total of 15 harmonics were used to adjust the shape of the orbital light curve. 

\subsubsection{V bandpass}

\begin{figure}
    \centering
    \includegraphics[width=1.0\columnwidth]{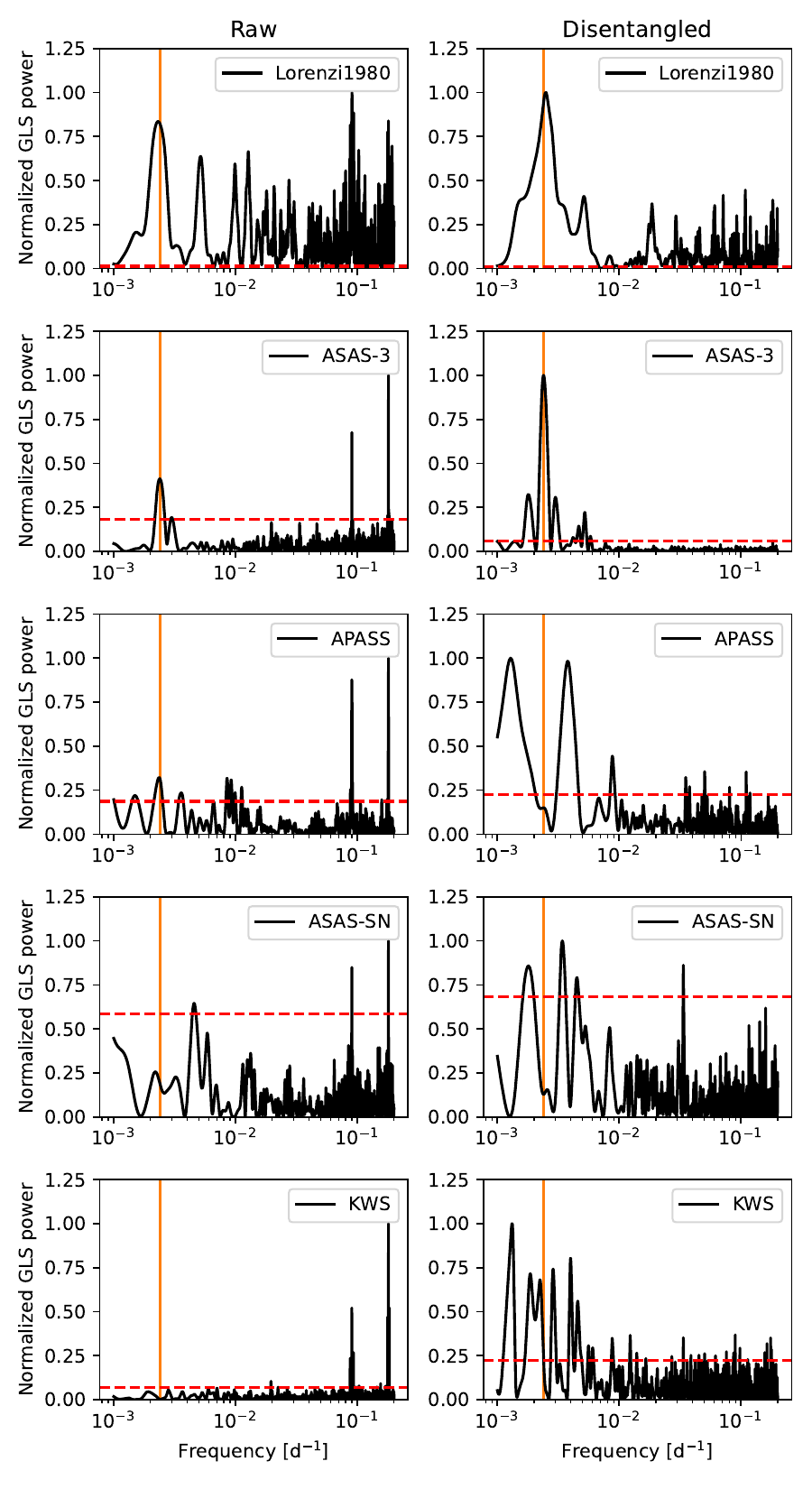}
    \caption{
    Normalised GLS periodograms for frequencies between 0.001 d$^{-1}$ (1\,000 d) and 0.2 d$^{-1}$ (5 d) of the raw data (left panels) and the disentangled data (right panels) for the Lorenzi1980, ASAS-3, APASS, ASAS-SN, and KWS datasets.
    In all panels, a FAP level of 99.9\% is indicated by horizontal dashed red lines, while the period of 417 days corresponding to the previously reported long cycle is marked with a vertical orange line.
    }
    \label{fig:photometry_gls}
\end{figure}

As discussed in Sect.~\ref{sec:orbital_cycle}, the GLS method is more suitable in the case of pure sinusoidal variations, such as the long cycle in AU Mon.
We construct GLS periodograms for the Lorenzi1980, ASAS-3, APASS, ASAS-SN, and KWS raw data, along with their corresponding disentangled light curves (see Fig.~\ref{fig:photometry_gls}).
A search for periodic signals between 5 and 1\,000 days was performed to test the effectiveness of the orbital period removal and to search for longer periods in the data.
We computed the false alarm probability (FAP) up to a 99.9\% confidence level to ensure that a peak is significant.
According to the FAP definition and the imposed confidence level, any peak above the FAP value has only a 0.1\% chance of being the result of random noise in the data \citep{VanderPlas2018}.
The FAP was calculated using the recommended bootstrap method in the GLS package, and is shown as a horizontal red line in each panel of Fig.~\ref{fig:photometry_gls}.

The strongest peaks observed in the raw periodograms correspond to the orbital period of 11.113 d (0.090000(14) d$^{-1}$) and its first harmonic (0.180000(15) d$^{-1}$).
In all cases, these peaks appear clean and isolated, with the exception of the Lorenzi1980 dataset, where the periodogram is noisier due to a higher observation cadence and greater point dispersion.
At lower frequencies, significant peaks around 0.002383 d$^{-1}$, which corresponds to the long cycle of $\sim417$ d (vertical orange line in Fig.~\ref{fig:photometry_gls}), are observed in the Lorenzi1980 and ASAS-3 datasets, and to a lesser extent in the APASS dataset.
Surprisingly, no significant peak at this frequency is observed in the power spectrum of the raw KWS or ASAS-SN data. For the latter, the most significant peak appears at a frequency of $\sim$0.0046 d$^{-1}$ ($\sim$217 d).

The lower panels of Fig.~\ref{fig:photometry_gls} show the GLS periodograms for the disentangled light curves, using the same period range as for the raw light curves.
Points showing a residual magnitude after disentangling greater than 0.5 mag were discarded. This prevents remnants of the orbital cycle, if present, from contaminating the periodogram through residual primary eclipse points.
The resulting periodograms show no significant peaks above the FAP level at higher frequencies, indicating the successful removal of the orbital period and its harmonics from the light curves.
The removal of the orbital period confirms the absence of the $\sim$417 day signal as the dominant periodicity in the light curves of APASS, ASAS-SN and KWS; instead, other periodicities dominate the periodogram.

In the APASS light curve, three prominent peaks appear at low frequency, corresponding to $\sim0.0013$ ($\sim769$), $\sim0.0038$ ($\sim263$), and $\sim0.0013$ ($\sim114$) d$^{-1}$ (d).
Similarly, the ASAS-SN periodogram shows three peaks at $\sim0.0018$ ($\sim556$), $\sim0.0034$ ($\sim294$), and $\sim0.0045$ ($\sim222$) d$^{-1}$ (d), with the strongest peak at $\sim0.0034$ d$^{-1}$.
Lastly, the KWS periodogram shows six peaks above the FAP level, corresponding to $\sim0.0013$ ($\sim769$), $\sim0.0018$ ($\sim556$), $\sim0.0022$ ($\sim454$), $\sim0.0029$ ($\sim344$), $\sim0.0040$ ($\sim250$), and $\sim0.0046$ ($\sim217$) d$^{-1}$ (d).
Some of the peaks are close to the frequencies corresponding to one and two years, indicating a likely aliasing origin resulting from the sampling of the data.
The three periodograms show significant peaks in the range of 250-290 d, which may be interpreted as the presence of a periodic signal with variations within this range.
It is also noticeable that a period of $\sim454$ d is present in the KWS data, which is close to the long cycle of $\sim417$ d, suggesting that some remnants of the long cycle may still be present in the data.

\begin{figure*}
    \centering
    \includegraphics[width=0.9\textwidth]{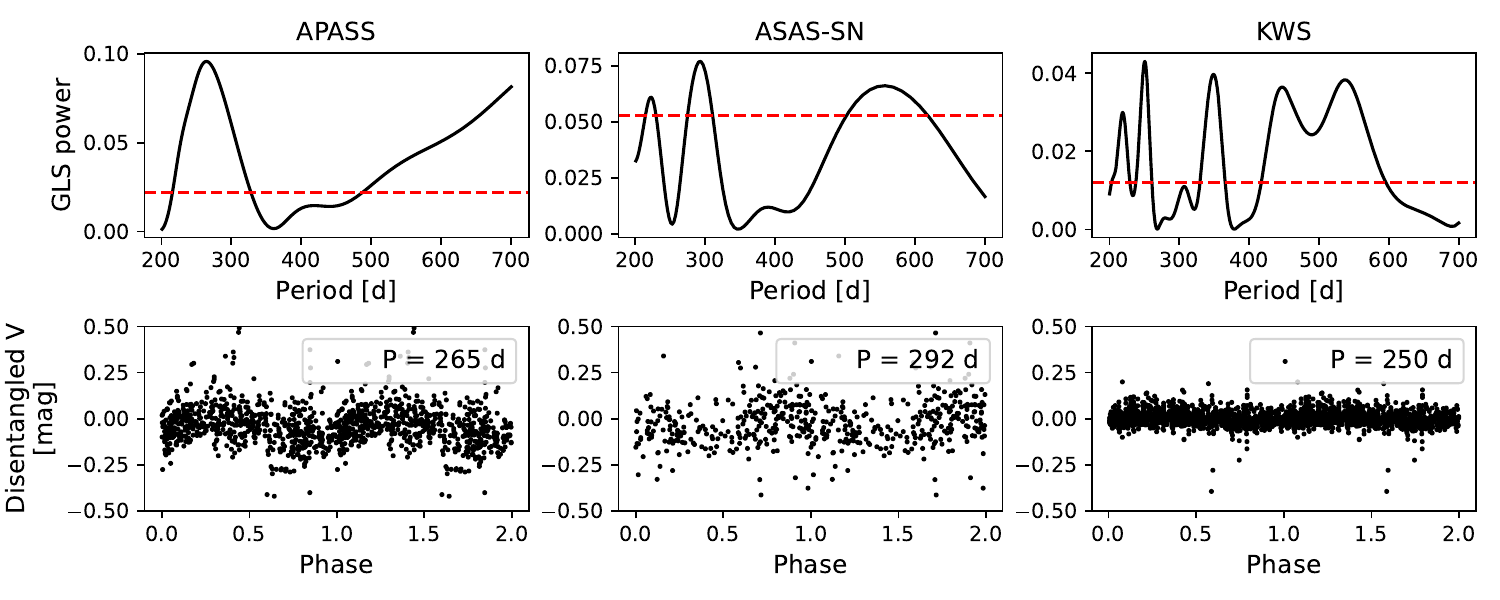}
    \caption{
    GLS periodograms for the APASS, ASAS-SN, and KWS datasets covering periods between 200 and 700 d (upper row), along with their respective folded light curves against the most significant peak within this range (bottom row).
    The 99.9\% FAP level is indicated by horizontal dashed red lines.
    }
    \label{fig:photometry_gls_zoom}
\end{figure*}

To better illustrate the behaviour of the periodicity in the range of 250-290 d, the periodograms were recalculated for the three datasets in a more constrained range of periods between 200 and 700 d.
The resulting periodograms are presented in Fig.~\ref{fig:photometry_gls_zoom}, where the periodogram of each dataset is shown together with its respective folded light curve for the period of greatest significance.
There are no appreciable differences between these constrained periodograms and the previous ones.
<however, the folded light curves exhibit a clear variability in the APASS dataset, which is less evident in the other two datasets.
In APASS, this variability has an amplitude of $\sim0.2$ mag, similar to that observed in ASAS-SN.
In contrast, the amplitude in KWS is considerably smaller, consistent with the disappearance observed in the case of the $\sim417$ d long period.

Additional confirmation of the disappearance of the long cycle comes from the analysis of the Weighted Wavelet Z (WWZ) transform \citep{Foster1996}.
We analysed the behaviour of the long cycle in the ASAS-3 and KWS light curves, after removing the orbital cycle, using the WWZ transform.
The combined light curve spans the interval $\sim52\,000-60\,000$ MJD.
The default value (0.001) for the decay parameter $\alpha$ was adopted to explore the period range between 200 and 700 days.
This value provides a good frequency resolution, allowing for clear identification of long-period signals such as the $\sim417$ day cycle, while also providing a smooth transition between datasets and minimising the presence of possible artefacts in the map.
The resolution of the time-period map was one day in the period domain and 200 time divisions in MJD, providing a sufficiently detailed analysis of the evolution of any periodicity within the period range.

\begin{figure}
    \centering
    \includegraphics[width=1.0\columnwidth]{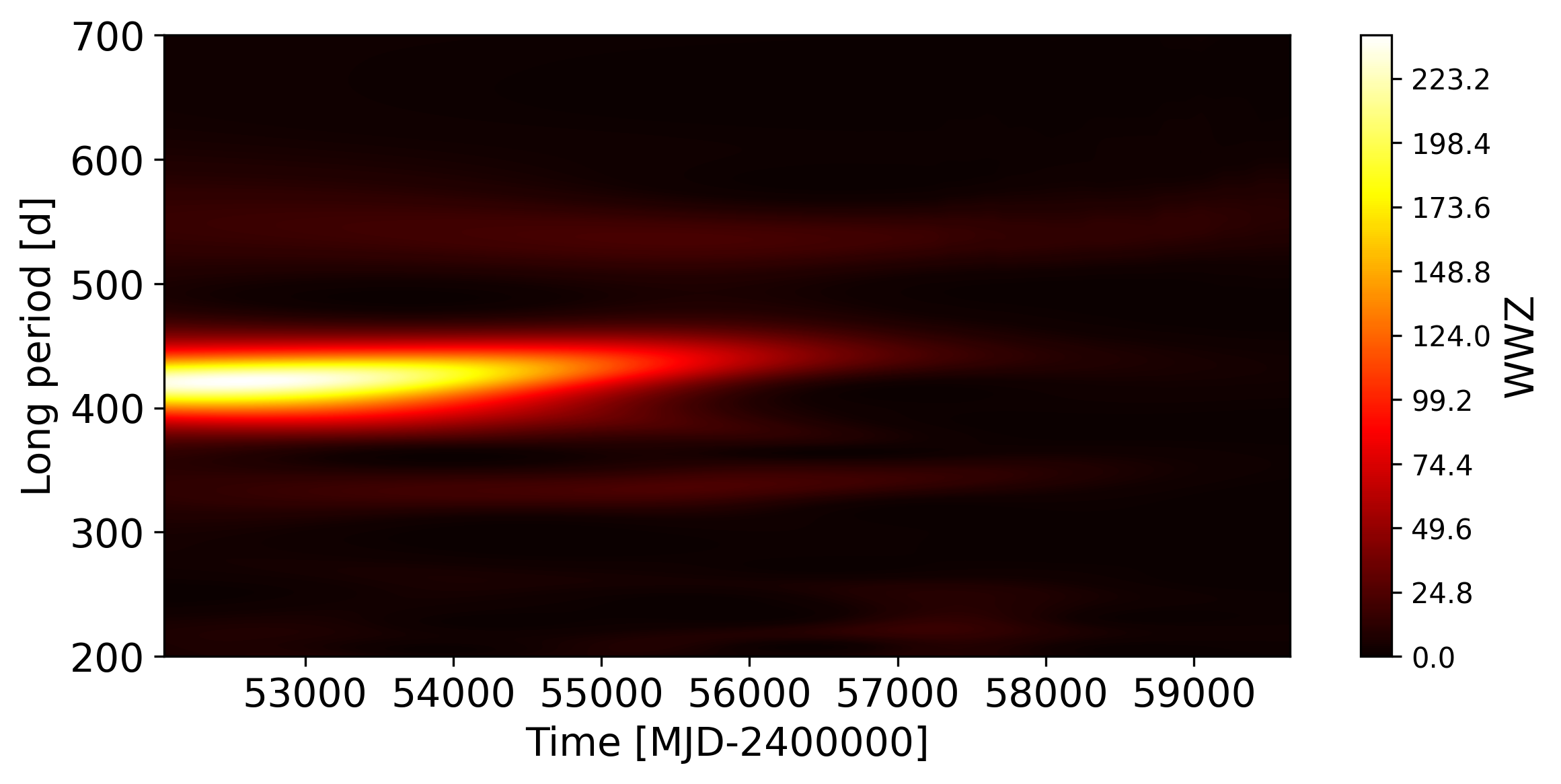}
    \caption{Weighted wavelength analysis of the disentangled light curves of ASAS-3 and KWS in the V filter. 
    The WWZ transform reveals a strong signal around 417 days that disappears after MJD $\sim$56\,000.}
    \label{fig:wwz-v}
\end{figure}

The resulting map is presented in Fig.~\ref{fig:wwz-v}.
In the ASAS-3 data, the long cycle signal of $\sim417$ days is clearly identifiable as a persistent and well-defined high-power band in the WWZ map, with power values exceeding 200.
However, at the time of the KWS photometry (MJD $\sim56,000$), the signal progressively weakened until it became visually undetectable, with WWZ values dropping below 50.
We interpret this continuous and sustained attenuation as evidence for the vanishing of the long cycle in AU Mon in the V filter.
In addition, the WWZ map shows persistent periodicities at $\sim350$ and 550 d, consistent with those observed in the periodograms of APASS, ASAS-SN, and KWS.

\subsubsection{Ic bandpass}

Previous studies have shown that the amplitude of the long cycle in DPVs is larger for redder bandpasses \citep{Mennickent2008, Michalska2009}.
Therefore, we expected the long cycle to be more evident in the Cousins I photometry available from the KWS survey.
To inspect the long-cycle behaviour at the Ic band, we applied the same methodology as described previously for the analysis of the V-band photometry; that is, we first removed the orbital period from the total light curve before analysing the disentangled photometry.

\begin{figure}
    \centering
    \includegraphics[width=1.0\columnwidth]{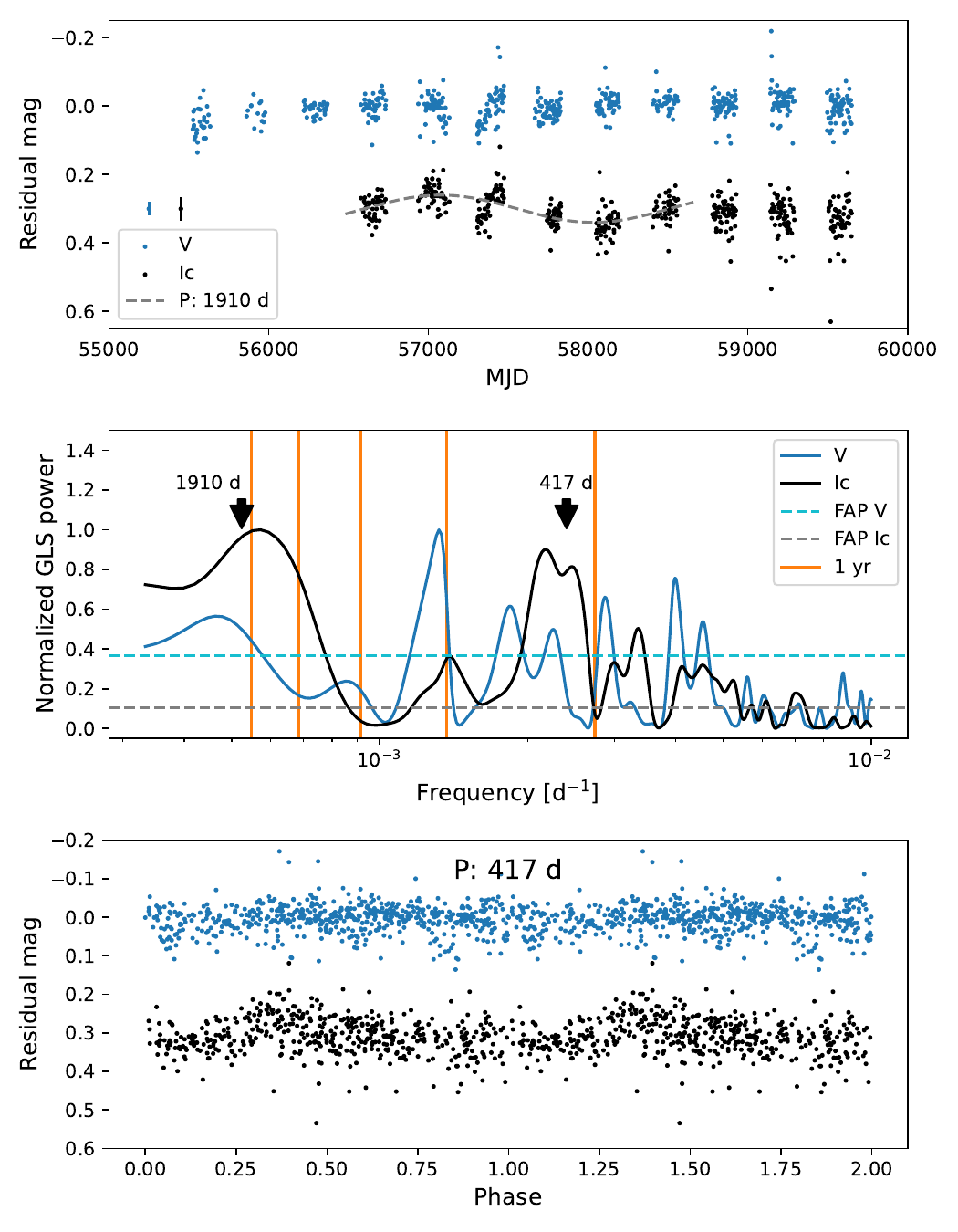}
    \caption{
    Light curves and periodograms of the disentangled KWS data in the V (blue) and Ic (black) bands.
    Upper panel: Disentangled light curves with the sinusoidal fit corresponding to a period of 1\,910 days shown as a dashed grey line.
    Average magnitude errors for both bands are presented on the left side of the plot.
    Middle panel: GLS periodograms of the disentangled light curves. The periodograms and their respective FAP limits are normalised to the observed maximum power.
    Frequencies corresponding to a one-year period and their first five harmonics are marked by orange lines.
    Arrows indicate frequencies corresponding to the periods of 417 and 1\,910 days.
    Bottom panel: Phased light curves corresponding to the period of 417 days.
    }
    \label{fig:photometry2}
\end{figure}

For comparison, the disentangled light curves in I and V bands are presented in the top panel of Fig.~\ref{fig:photometry2}. The residual magnitudes corresponding to the V filter (blue points) do not show any evidence of long-cycle variations, as established in the previous subsection.
This is not the case for the light curve of the Ic bandpass, for which the data follow a sinusoidal pattern with an amplitude of $\sim$0.05\,mag. Intriguingly, this variability appears to suddenly stop after the MJD $\sim$58\,800.
We performed a sinusoidal fit to the data in this band, considering only those points prior to the aforementioned MJD. The best sinusoidal fit indicates a variation with an amplitude of 0.04(3) mag and a period of 1910(70) d (dashed grey line in the figure). The ephemeris for this long period follows:

\begin{equation}
T_\mathrm{max} = \mathrm{MJD} \, 57\,195(6) + 1\,910(70) \times E.
\end{equation}

The periodograms of both light curves (V and Ic) are presented in the middle panel of Fig.~\ref{fig:photometry2}. Both periodograms were constructed using the GLS method, with a period range between 100 and 3\,000 days. The upper limit was selected as it is close to the time range of the Ic observations ($\sim$3\,070 d). The FAP levels at 99.9\% were calculated and are shown as horizontal dashed lines. For clarity, both the GLS power and the FAP level of each periodogram were normalised by their respective GLS power maxima.

In contrast to the V band, the data in the Ic band show the most prominent peak at $\sim1\,785$ days, with the secondary peaks close to $\sim$460 and $\sim$406 days. The first peak is roughly consistent with the period of $\sim1\,910$ days determined by the sinusoidal fit, while the second and third peaks are close to the previously reported period of $\sim$417 days, possibly representing remnants of the long cycle.
An additional peak is observed at $\sim$720 days; however, similar to the case of the V dataset, this peak is likely an artefact resulting from the one-year observations schedule (marked with orange lines in the middle panel of Fig.\ref{fig:photometry2}).

A remnant of the $\sim417$ days long cycle can still be observed in the KWS data. The bottom panel of Fig.~\ref{fig:photometry2} shows the data phased against this period using the ephemeris published by D10. In both cases, a small variation in magnitude can be seen as a function of phase, which is more evident in the Ic light curve than in the V band.

\subsection{APASS multi-wavelength analysis}

The data provided by APASS consist of several filters covering the optical spectrum. 
The effective wavelengths of these filters are presented in Table~\ref{tab:summary_photometry}.
This allowed us to perform a multi-wavelength analysis of the system's light curve.
The observations span different time ranges for each filter; however, we identified a period of significant overlap. Thus, we restricted this analysis to the observations carried out within this overlap, corresponding to epochs between MJD 56\,150 and 56\,850, for which several hundred data points are available for each filter: 457 for the B and V, 346 for g, 341 for r, 343 for i, and 372 for Z.
The time coverage of these datasets allows us to analyse both the orbital period and any other long-period modulations.

\begin{figure}
\centering
\includegraphics[width=0.95\columnwidth]{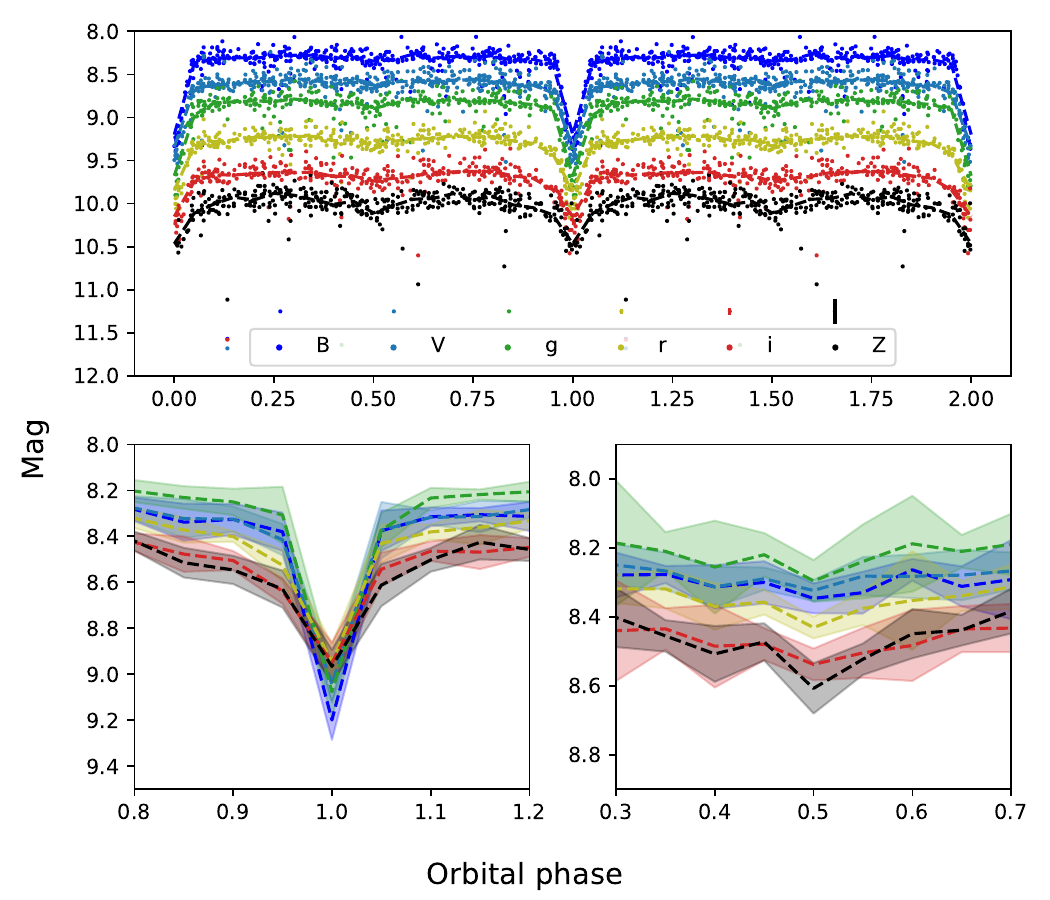}
\caption{APASS multi-wavelength analysis of the orbital period.
Top: APASS light curves in the filters (from top to bottom): B, V, g, r, i, and Z. A shift in magnitude has been applied to every light curve for better visualisation.
Bottom: Smoothed light curves without magnitude shift around the primary and secondary eclipse.}
\label{fig:APASS_multiwavelength_1}
\end{figure}

The analysis of the orbital cycle is presented in Fig.~\ref{fig:APASS_multiwavelength_1}. The light curve for each filter was phased against the orbital period and ephemeris reported by D10. The resulting phased light curves are presented in the top panel of the figure, where different colours indicate the different data points corresponding to each filter.
The data points were grouped in bins of 0.05 phase width, and their corresponding magnitudes were averaged to produce a smoothed version of the light curve (dashed lines in the top panel of Fig.~\ref{fig:APASS_multiwavelength_1}, with the colour limits indicating the standard deviation at each bin). 

All the light curves show a prominent primary eclipse, with the depth being greatest for the bluer filters. The lower panels of Fig.~\ref{fig:APASS_multiwavelength_1} show a zoom of the primary (left panel) and secondary (right panel) eclipses, using only the smoothed light curves.
Unlike the upper panel, the magnitudes are not shifted, so it can be seen that the system is brighter in the g band. The depth of the primary eclipse is strongest in the B filter and weakest in the Z filter, while for the secondary eclipse, the opposite behaviour is observed.
To quantify the depth of the primary and secondary eclipse for each filter, the average magnitudes, $m_\mathrm{pri}$ and $m_\mathrm{sec}$, for the primary and secondary eclipses, respectively, were obtained.
They were compared with the mean magnitude observed at the quarter phases of the orbital cycle (orbital phases 0.25 and 0.75), $m_\mathrm{qua}$.
The difference between $m_\mathrm{qua}$ and $m_\mathrm{pri}$ or $m_\mathrm{sec}$ can be used to quantify the depth of each eclipse.
The results are presented in Table~\ref{tab:apass_magnitudes}, where the mean magnitudes are given together with their respective standard deviations, presented as uncertainties.
The measurements corroborate the visual analysis previously presented, with the primary eclipse being deeper for the bluer filters.
This is expected, as the most luminous component in the system is the hot component \citep{Djurasevic2010}, which dominates the blue part of the spectrum. When eclipsed, the decrease in brightness is strongest for the B filter ($\Delta$B = 0.76$\pm$0.25 mag), which has the bluest $\lambda_\mathrm{eff}=442$ nm (Table~\ref{tab:summary_photometry}).
For the reddest filter, the Z filter ($\lambda_\mathrm{eff}=876$ nm), the contribution of the hot component is smaller, causing shallower primary eclipses ($\Delta$Z = 0.39$\pm$0.4).
For the same reason, the secondary eclipse is more evident in this filter, as the contribution of the secondary star is proportionally higher ($\Delta$Z = 0.07$\pm$0.39).
It should be noted that the significant dispersion in the data, particularly in the redder filters, prevents a more precise measurement of the eclipses.

\begin{table}[]
    \centering
    \caption{Measured mean magnitudes and standard deviations at the primary and secondary eclipses, as well as at the quarter phases, for the different APASS datasets.}
    \label{tab:apass_magnitudes}
    \begin{tabular}{lrrr}
    \hline\hline
    Filter & $m_\mathrm{qua}$ & $m_\mathrm{pri}$ & $m_\mathrm{sec}$ \\
    \hline\noalign{\smallskip}
    B & 8.32$\pm$0.11 & 9.08$\pm$0.23 & 8.32$\pm$0.09 \\
    V & 8.29$\pm$0.15 & 8.98$\pm$0.19 & 8.32$\pm$0.12 \\
    g & 8.24$\pm$0.14 & 8.98$\pm$0.24 & 8.28$\pm$0.11 \\
    r & 8.40$\pm$0.67 & 8.94$\pm$0.17 & 8.39$\pm$0.07 \\
    i & 8.48$\pm$0.44 & 8.94$\pm$0.16 & 8.52$\pm$0.09 \\
    Z & 8.48$\pm$0.38 & 8.87$\pm$0.15 & 8.55$\pm$0.09 \\
    \hline
    \end{tabular}
\end{table}

\begin{figure}
\centering
\includegraphics[width=0.95\columnwidth]{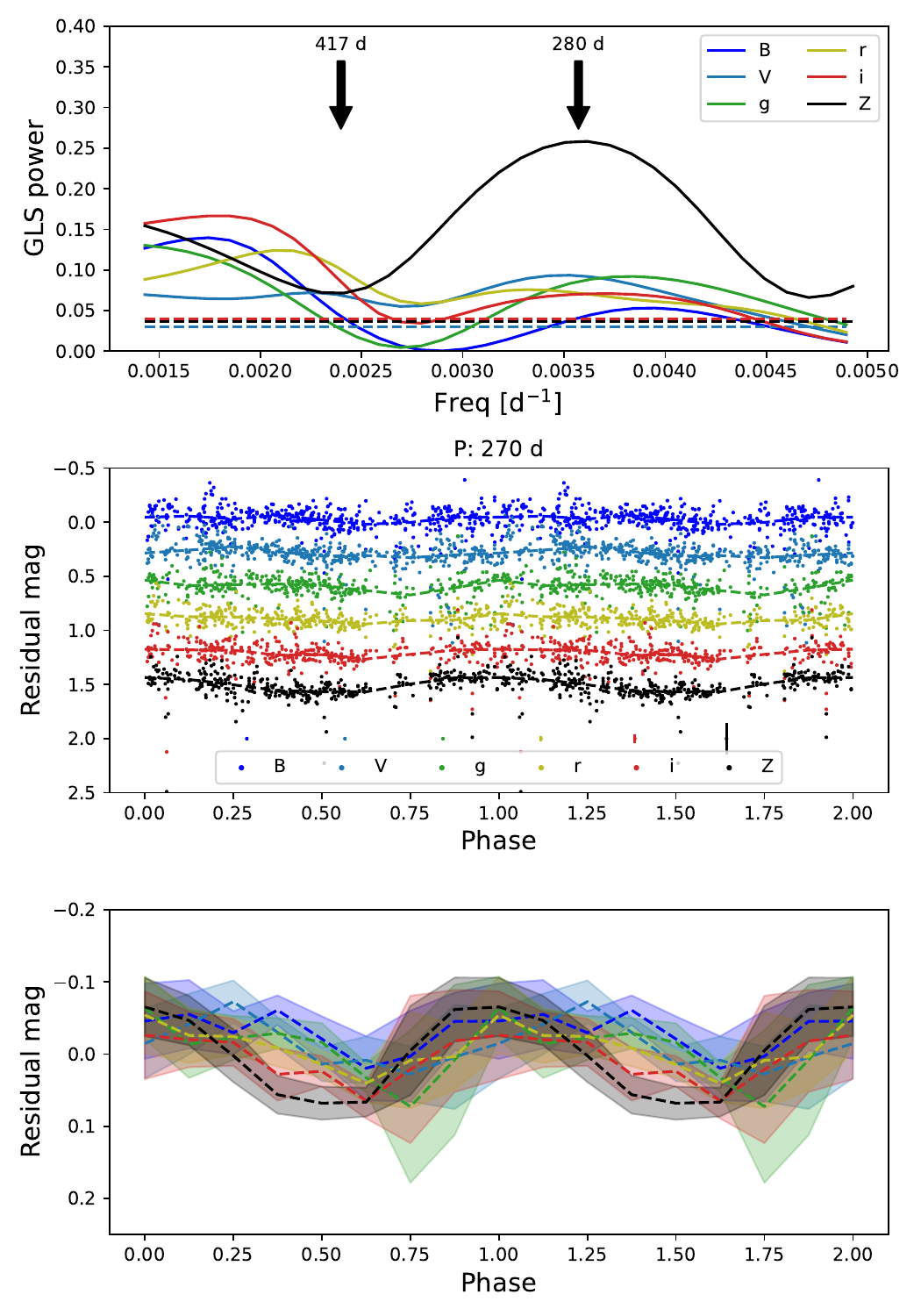}
\caption{APASS multi-wavelength analysis of the long cycle.
Top: Periodogram for the different datasets.
Middle: Phased light curves with a period of 280 d.
Bottom: Smoothed light curves.}
\label{fig:APASS_multiwavelength_2}
\end{figure}

For the analysis of the possible long-cycle modulations, we disentangled the orbital frequencies from the light curves using the same methodology as described in Section~\ref{sec:long_cycle}. The periodogram for the disentangled light curves is presented in the top panel of Fig.~\ref{fig:APASS_multiwavelength_2}. We searched for periodicities between 100 and 700 days, finding the strongest signal at $\sim$280 days, with the second strongest peak located at $\sim$175 days. These signals are strongest for the Z filter, consistent with the previous statement that the long-cycle modulations are stronger in the redder filters.
No signal of the 417\,d long cycle is present in the periodogram.

The light curves phased to the strongest periodicity of 280 days are presented in the middle panel of Fig.~\ref{fig:APASS_multiwavelength_2}, together with their smoothed counterparts in the bottom panel. Here, sinusoidal modulation appears clearly visible in the Z filter with an amplitude $\Delta$Z=0.14$\pm$0.02 mag.
In the rest of the filters, the modulation is less evident.
In the bluer filters (B and V), the maximum appears shifted with respect to the Z filter. 
With the current data, we cannot determine whether this shift is a real feature of the light curves or an artefact produced by the high dispersion observed in the data.

\section{Discussion}

\subsection{Stability of the orbital period}
Our work indicates that the orbital period of AU Mon has remained constant over the last century within the given uncertainties (Table~\ref{tab:orbital_period_measurements}). These results support the previously established stability of the orbital period \citep{Kreiner2004}.
As pointed out by \citet{Mennickent2014}, the stability of the orbital period should imply the presence of a mechanism for the mass and angular loss of the system.
The fractional change of the orbital period, $\dot{P}_\mathrm{O}/P_\mathrm{O}$, in a binary system under a non-conservative mass transfer regime can be expressed as

\begin{equation}
    \frac{\dot{P}_\mathrm{o}}{P_\mathrm{o}} = -\frac{\dot{M}_2}{M_2}\,[3 - 3\beta\frac{M_2}{M_1} - (1-\beta)\frac{M_2}{M_\mathrm{T}} - 3(1-\beta)\alpha\frac{M_1}{M_\mathrm{T}}] + 3\frac{\dot{J}}{J},
    \label{eqn:dotP}
\end{equation}

where $M_1$ and $M_2$ correspond to the mass of the gainer and the donor, respectively, and $M_\mathrm{T}$ to the total mass of the system (i.e. $M_1+M_2$) \citep{Verbunt1993ARA&A..31...93V}. The parameters $\beta$ and $\alpha$ are the fraction of the transferred mass that is accreted into the gainer star, and the fraction of specific angular momentum with respect to the donor star that is carried away by the mass loss from the system, respectively, while $\dot{J}/J$ indicates the fractional change of angular momentum by other sources (e.g. magnetic braking). For simplicity, we ignored the last term in our analysis.

If the donor star is less massive than the gainer, the value of $\dot{P}_\mathrm{o}$ will always be positive, regardless of the values of $\beta$ and $\alpha$. Considering a value of 7.6$\times$10$^{-6}$ M$_\odot$\,yr$^{-1}$ for $\dot{M}$ \citep{Mennickent2014} and the component masses in the system \citep{Djurasevic2010}, even under the most favourable (and unrealistic) conditions for mass and angular momentum loss from the system (i.e. $\beta$=0, $\alpha$=1), a minimum value for $\dot{P}_{\mathrm{o}}$ of $\sim$1.8 s yr$^{-1}$ is reached.
This value is considerably higher than the 0.038$\pm$0.04 s\,yr$^{-1}$ derived from our O-C analysis. This indicates that the $\dot{M}$ value provided by \citet{Mennickent2014} is significantly overestimated.
This result is similar to the case of OGLE-LMC-ECL-14413, for which a value of $\dot{M}$ too high to be consistent with the observed orbital period stability was obtained \citep{Mennickent+2025A&A...693A.217M}. This suggests a systematic overestimation of the $\dot{M}$ values derived from evolutionary models and/or a significant loss of matter and angular momentum from these systems.

By contrast, \citet{Atwood2012} provided a lower limit for $\dot{M}$ of $\sim2.4\times10^{-9}$ M$_\odot$\,yr$^{-1}$. This lower limit leads to $\dot{P}_\mathrm{o}\sim5\times10^{-3}$ s\,yr$^{-1}$ in the case of fully conservative mass transfer ($\beta=1$, $\alpha=0$), which is consistent with our derived value of $0.038\pm0.040$ s\,yr$^{-1}$.
Similarly, by assuming a fully conservative mass transfer, our results yield to $\dot{P}_\mathrm{o} \sim 2\times10^{-8}$ M$_\odot$\,yr$^{-1}$.
If we instead consider a case of non-conservative mass transfer, the derived value of $\dot{P}_\mathrm{o}$ from Equation~\ref{eqn:dotP} will be smaller, and thus allows a higher value of $\dot{M}$. In the most extreme case (i.e. $\beta$=0, $\alpha$=1) we derive $\dot{M} \sim1.5\times10^{-7}$ M$_\odot$\,yr$^{-1}$ which is still compatible with $\dot{P}_\mathrm{o}$ derived from the O-C analysis.
Therefore, we conclude that the $\dot{M}$ value in AU Mon must lie between $\sim2\times10^{-8}$ and $\sim1.5\times10^{-7}$ M$_\odot$\,yr$^{-1}$.
If there is a significant loss of angular momentum in the binary due to other sources, the increase in $\dot{P}_\mathrm{o}$ will be compensated, thereby permitting higher values of $\dot{M}$.

\subsection{Origin of the long-cycle variation}

While the origin of the long cycle in DPV systems remains uncertain, some studies have proposed physical mechanisms that could explain its nature. In particular, \citet{Schleicher2017} suggested that this phenomenon may be related to the presence of an active magnetic dynamo in the donor star, which cyclically modulates its quadrupole moment, following the mechanism proposed by \citet{Applegate1992}. These changes in the quadrupole moment would induce periodic variations in the donor's equatorial radius, thereby affecting the mass transfer rate onto the companion star. This modulation could have direct consequences for the structure of the accretion disc, leading to changes in its temperature, radial extension, and luminosity, as observed in systems such as OGLE-LMC-DPV-097 \citep{Garces2018}, OGLE-BLG-ECL-157529 \citep{Mennickent+2020A&A...641A..91M}, and OGLE-LMC-ECL-14413 \citep{Mennickent+2025A&A...693A.217M}. In this context, the long cycle would be the observable outcome of a complex interaction between the donor's magnetic activity, the dynamics of the mass flow, and the structural response of the disc.

The H$\alpha$ profile of AU Mon shows a clear double peak, characteristic of an accretion disc. However, these profiles exhibit variability modulated by the long cycle, with a deeper central absorption observed during the low phase of the cycle \citep{Barria&Mennickent2011, Atwood2012}. These changes could be associated with structural variations in the accretion disc, as suggested by the multi-wavelength spectroscopic study of \citet{Armeni_2022}. Nevertheless, \citet{Djurasevic2010} argued that changes in the disc structure alone cannot fully explain the long cycle. It is worth noting that their analysis was based on V-filter photometry, whereas the most pronounced changes in the orbital light-curve morphology in other DPV systems have been observed in the I band.

Therefore, although it is plausible that a mechanism similar to those observed in the aforementioned DPVs is operating in AU Mon, our current data and analysis do not allow us to confirm changes in its orbital light curve associated with the long cycle, which could be interpreted as a result of structural variations in the accretion disc.
Additional information may arise from multi-wavelength photometry of the long cycle. Our analysis of the APASS multi-wavelength photometry reveals a possible shift in the peak of the $\sim290$ d long period between the B and Z filters. Such an analysis has not been carried out on other DPVs and presents an interesting opportunity to reveal the nature of the long cycle.

There are special cases of DPVs that show a significant decrease in their long periods, such as OGLE 05155332-6925581 \citep{Mennickent2008} and OGLE-LMC-DPV-065 \citep{Mennickent+2019MNRAS.487.4169M}. These systems show a substantial decrease ($\sim$20\%) in the period of their long cycle but no changes in their amplitude.
So far, only in OGLE-BLG-ECL-157529 has a simultaneous decrease in the long period and long-cycle amplitude been observed.
What we observe in AU Mon, however, is different. We observe a vanishing of the amplitude of the long cycle, rather than a decrease in the long period. The gradual dissapearence of the long cycle is reflected in the consistent weakening over time of the frequency associated with the $\sim$417\,d signal in the V filter from the Lorenzi1980 data ($\sim$1975-1979), the ASAS-3 data ($\sim$2001-2009), and the KWS data ($\sim$2010-2020), although some remnants of this period can still be observed in the Ic filter from KWS.
More puzzling, additional periodicities (at $\sim280$ and $\sim1\,910$ days) have appeared temporarily in the redder filters, but these also subsequently disappeared.
The observed behaviour in AU Mon is more similar to the short disappearance of the long cycle reported in TYC 5353-1137-1 \citep{Rosales&Mennickent2018IBVS.6248....1R}. In this DPV, the long cycle suddenly vanished for $\sim1\,500$\,d before reappearing at the same period and amplitude as before. Although the study of TYC 5353-1137-1 spanned only $\sim10$ years and only considered the V filter, it opens the possibility that the long cycle in AU Mon may reappear. Future observations will clarify whether this is the case or not.

A recent study of the evolutionary path of the galactic DPV V4142 Sgr shows that the DPV phenomenon may originate at different phases of the evolutionary path of the binary system \citep{Rosales+2024A&A...689A.154R}. The authors conclude that the long cycle, modelled under the \citet{Schleicher2017} prescription, may appear after the mass ratio inversion and last until the current age of the system.
Additionally, \citet{Rosales+2024A&A...689A.154R} identified a compact region in the HR diagram where the hot components of several DPVs, including AU Mon, are located. This indicates a common evolutionary stage for these systems, from which we can correlate the behaviour of the long cycle.
Under the Applegate mechanism, our main result--the vanishing of the long cycle in AU Mon--would imply changes in the parameters governing the aforementioned mechanism that rendered the appearance of the long cycle inviable.
If so, this would imply drastic changes in the internal structure of the cool star in AU Mon.
Thus, the vanishing of the long cycle is a strong constraint for any competent model and should be analysed in detail in future theoretical studies.

\section{Conclusions}

We analysed a series of archival photometric data of the DVP system AU Mon, covering 46.3 years, with the main goal of studying the stability of the orbital period and the long cycle during this time.
Employing time-series analysis, we derived an orbital period variation $\dot{P}_\mathrm{o} = 0.038\pm0.040$ s\,yr$^{-1}$, consistent with a constant orbital period within the uncertainties.
We placed constraints on the mass transfer rate of the system by considering both fully conservative and fully non-conservative mass transfer regimes. This led to a $\dot{M}$ between $2\times10^{-8}$ and $1.5\times10^{-7}$ M$_\odot$\,yr$^{-1}$.

In addition, we corroborated the existence of the previously reported $\sim$417\,d long cycle in the V filter but also confirmed its disappearance during the most recent years of observations. 
This corresponds to the second reported disappearance of the long cycle in a DPV, after the case of TYC 5353-1137-1.
The analysis of the light curves in the Ic and Z filters, on the other hand, revealed the appearance of short-lived periodicities (at $\sim$280 and $\sim$1910\,d).
The timing of these periodicities coincides with the lack of the $\sim$417\,d long cycle, and they were detected for a limited time before also vanishing.

These findings illustrate the complex nature of the long-term photometric behaviour of AU Mon and potentially other DPV systems.
This constitutes a strong constraint for any competent model aimed at explaining the cyclic long-term variability, particularly those invoking a magnetic dynamo in the donor star.
It also highlights the importance of continuing long-term photometric monitoring of this binary system, at several time scales, to understand the real nature of its variability.

\section*{Data availability}
The photometric data used in this work are publicly available from the web pages of the different sources, at Zenodo\footnote{\href{https://zenodo.org/records/15178889}{AU Mon photometric data}}, or upon reasonable request to the author.

\begin{acknowledgements}

The authors thank the anonymous referee for their valuable comments and positive feedback, which significantly improved the paper's quality and readability. 
L.C. acknowledges support from ANID-Subdireccion de capital humano/doctorado nacional/2022-21220607. 
R.E.M. acknowledges support by the ANID BASAL project Centro de Astrof{\'{i}}sica y Tecnolog{\'{i}}as Afines ACE210002 (CATA).
D.B. acknowledges support by Fondecyt grant Nro. 11230261.
Part of this work was supported by the German \emph{Deut\-sche For\-schungs\-ge\-mein\-schaft, DFG\/} project number Ts~17/2--1.
J.G. acknowledges support by ANID grant Nro. 21202285.
M.J. acknowledges support from the Ministry of Science, Technological Development and Innovation, grant Nro. 451-03-136/2025-03/200002.

\end{acknowledgements}

\bibliographystyle{aa}
\bibliography{aa55156-25.bib}

\begin{thebibliography}{62}
\expandafter\ifx\csname natexlab\endcsname\relax\def\natexlab#1{#1}\fi

\bibitem[{Hip(1997)}]{Hipparcos97}
 1997, ESA Special Publication, Vol. 1200, {The HIPPARCOS and TYCHO catalogues.
  Astrometric and photometric star catalogues derived from the ESA HIPPARCOS
  Space Astrometry Mission}

\bibitem[{{Applegate}(1992)}]{Applegate1992}
{Applegate}, J.~H. 1992, \apj, 385, 621

\bibitem[{{Armeni} \& {Shore}(2022)}]{Armeni_2022}
{Armeni}, A. \& {Shore}, S.~N. 2022, \aap, 664, A103

\bibitem[{{Atwood-Stone} {et~al.}(2012){Atwood-Stone}, {Miller}, {Richards},
  {Budaj}, \& {Peters}}]{Atwood2012}
{Atwood-Stone}, C., {Miller}, B.~P., {Richards}, M.~T., {Budaj}, J., \&
  {Peters}, G.~J. 2012, \apj, 760, 134

\bibitem[{{Barr{\'\i}a} \& {Mennickent}(2011)}]{Barria&Mennickent2011}
{Barr{\'\i}a}, D. \& {Mennickent}, R.~E. 2011, in Astronomical Society of the
  Pacific Conference Series, Vol. 447, Evolution of Compact Binaries, ed.
  L.~{Schmidtobreick}, M.~R. {Schreiber}, \& C.~{Tappert}, 263

\bibitem[{{Barr{\'\i}a} {et~al.}(2013){Barr{\'\i}a}, {Mennickent},
  {Schmidtobreick}, {Djura{\v{s}}evi{\'c}}, {Ko{\l}aczkowski}, {Michalska},
  {Vu{\v{c}}kovi{\'c}}, \& {Niemczura}}]{Barria2013}
{Barr{\'\i}a}, D., {Mennickent}, R.~E., {Schmidtobreick}, L., {et~al.} 2013,
  \aap, 552, A63

\bibitem[{{Cerruti-Sola} \& {Lorenzi}(1977)}]{Cerruti-Sola&Lorenzi1977}
{Cerruti-Sola}, M. \& {Lorenzi}, L. 1977, Information Bulletin on Variable
  Stars, 1348, 1

\bibitem[{{Desmet} {et~al.}(2010){Desmet}, {Fr{\'e}mat}, {Baudin}, {Harmanec},
  {Lampens}, {Pacheco}, {Briquet}, {Degroote}, {Neiner}, {Mathias}, {Poretti},
  {Rainer}, {Uytterhoeven}, {Amado}, {Valtier}, {Pr{\v{s}}a}, {Maceroni}, \&
  {Aerts}}]{Desmet2010}
{Desmet}, M., {Fr{\'e}mat}, Y., {Baudin}, F., {et~al.} 2010, \mnras, 401, 418

\bibitem[{{Djura{\v{s}}evi{\'c}} {et~al.}(2010){Djura{\v{s}}evi{\'c}},
  {Latkovi{\'c}}, {Vince}, \& {Cs{\'e}ki}}]{Djurasevic2010}
{Djura{\v{s}}evi{\'c}}, G., {Latkovi{\'c}}, O., {Vince}, I., \& {Cs{\'e}ki}, A.
  2010, \mnras, 409, 329

\bibitem[{{Floria}(1937)}]{Floria1937}
{Floria}, N. 1937, Trudy Gosudarstvennogo Astronomicheskogo Instituta, 8, 5

\bibitem[{{Foster}(1996)}]{Foster1996}
{Foster}, G. 1996, \aj, 112, 1709

\bibitem[{{Garc{\'e}s} {et~al.}(2019){Garc{\'e}s}, {Mennickent},
  {Djura{\v{s}}evi{\'c}}, \& {Poleski}}]{Garces2019}
{Garc{\'e}s}, J., {Mennickent}, R.~E., {Djura{\v{s}}evi{\'c}}, G., \&
  {Poleski}, R. 2019, Contributions of the Astronomical Observatory Skalnate
  Pleso, 49, 355

\bibitem[{{Garc{\'e}s L} {et~al.}(2018){Garc{\'e}s L}, {Mennickent},
  {Djura{\v{s}}evi{\'c}}, {Poleski}, \& {Soszy{\'n}ski}}]{Garces2018}
{Garc{\'e}s L}, J., {Mennickent}, R.~E., {Djura{\v{s}}evi{\'c}}, G., {Poleski},
  R., \& {Soszy{\'n}ski}, I. 2018, \mnras, 477, L11

\bibitem[{{Garrido} {et~al.}(2013){Garrido}, {Mennickent},
  {Djura{\v{s}}evi{\'c}}, {Ko{\l}aczkowski}, {Niemczura}, \&
  {Mennekens}}]{Garrido2013}
{Garrido}, H.~E., {Mennickent}, R.~E., {Djura{\v{s}}evi{\'c}}, G., {et~al.}
  2013, \mnras, 428, 1594

\bibitem[{{Glazunova} {et~al.}(2008){Glazunova}, {Yushchenko}, {Tsymbal},
  {Mkrtichian}, {Lee}, {Kang}, {Valyavin}, \&
  {Lee}}]{Glazunova+2008AJ....136.1736G}
{Glazunova}, L.~V., {Yushchenko}, A.~V., {Tsymbal}, V.~V., {et~al.} 2008, \aj,
  136, 1736

\bibitem[{{G{\l}owacki} {et~al.}(2025){G{\l}owacki}, {Soszy{\'n}ski},
  {Udalski}, {Szyma{\'n}ski}, {Skowron}, {Skowron}, {Mr{\'o}z}, {Pietrukowicz},
  {Poleski}, {Koz{\l}owski}, {Iwanek}, {Wrona}, {Ulaczyk}, {Rybicki},
  {Gromadzki}, {Mr{\'o}z}, \& {Urbanowicz}}]{Glowacki+2025arXiv250315596G}
{G{\l}owacki}, M., {Soszy{\'n}ski}, I., {Udalski}, A., {et~al.} 2025, arXiv
  e-prints, arXiv:2503.15596

\bibitem[{{Haas} {et~al.}(2012){Haas}, {Hackstein}, {Ramolla}, {Drass},
  {Watermann}, {Lemke}, \& {Chini}}]{Haas+2012}
{Haas}, M., {Hackstein}, M., {Ramolla}, M., {et~al.} 2012, Astronomische
  Nachrichten, 333, 706

\bibitem[{{Hackstein} {et~al.}(2015){Hackstein}, {Fein}, {Haas}, {Ramolla},
  {Pozo Nu{\~n}ez}, {Barr Dom{\'\i}nguez}, {Kaderhandt}, {Thomsch},
  {Niedworok}, {Westhues}, \& {Chini}}]{Haas+2015}
{Hackstein}, M., {Fein}, C., {Haas}, M., {et~al.} 2015, Astronomische
  Nachrichten, 336, 590

\bibitem[{{Henden}(2011)}]{Henden2011}
{Henden}, A. 2011, \skytel, 121, 34

\bibitem[{{Hoffmeister}(1931)}]{Hoffmeister1931}
{Hoffmeister}, C. 1931, Astronomische Nachrichten, 242, 129

\bibitem[{{Kalimeris} {et~al.}(1994){Kalimeris}, {Rovithis-Livaniou}, \&
  {Rovithis}}]{Kalimeris94}
{Kalimeris}, A., {Rovithis-Livaniou}, H., \& {Rovithis}, P. 1994, \aap, 282,
  775

\bibitem[{{Kreiner}(2004)}]{Kreiner2004}
{Kreiner}, J.~M. 2004, \actaa, 54, 207

\bibitem[{{Lause}(1949)}]{Lause1949}
{Lause}, F. 1949, Astronomische Nachrichten, 277, 40

\bibitem[{{Lorenzi}(1980)}]{Lorenzi80}
{Lorenzi}, L. 1980, \aaps, 40, 271

\bibitem[{{Lorenzi}(1985)}]{Lorenzi1985}
{Lorenzi}, L. 1985, Information Bulletin on Variable Stars, 2704, 1

\bibitem[{{Maehara}(2015)}]{Maehara2015_KWS}
{Maehara}, H. 2015, {Journal of Space Science Informatics Japan}

\bibitem[{{Mennickent}(2014)}]{Mennickent2014}
{Mennickent}, R.~E. 2014, \pasp, 126, 821

\bibitem[{{Mennickent}(2017)}]{Mennickent2017}
{Mennickent}, R.~E. 2017, Serbian Astronomical Journal, 194, 1

\bibitem[{{Mennickent}(2022)}]{Mennickent2022}
{Mennickent}, R.~E. 2022, Galaxies, 10, 15

\bibitem[{{Mennickent} {et~al.}(2004){Mennickent}, {Assmann}, G., \&
  W.}]{Mennickent2004}
{Mennickent}, R.~E., {Assmann}, P., G., P., \& W., G. 2004, ASP Conf. Ser.,
  Proc. Workshop The Light-Time Effect in Astrophysics. Astron. Soc. Pac., San
  Francisco, Vol 335, p. 129

\bibitem[{{Mennickent} {et~al.}(2019){Mennickent}, {Cabezas},
  {Djura{\v{s}}evi{\'c}}, {}, {Rivinius}, {Hadrava}, {Poleski},
  {Soszy{\'n}ski}, {Celed{\'o}n}, {Astudillo-Defru}, {Raj},
  {Fern{\'a}ndez-Trincado}, {Schmidtobreick}, {Tappert}, {Neustroev}, \&
  {Porritt}}]{Mennickent+2019MNRAS.487.4169M}
{Mennickent}, R.~E., {Cabezas}, M., {Djura{\v{s}}evi{\'c}}, {et~al.} 2019,
  \mnras, 487, 4169

\bibitem[{{Mennickent} {et~al.}(2005){Mennickent}, {Cidale}, {D{\'\i}az},
  {Pietrzy{\'n}ski}, {Gieren}, \& {Sabogal}}]{Mennickent2005}
{Mennickent}, R.~E., {Cidale}, L., {D{\'\i}az}, M., {et~al.} 2005, \mnras, 357,
  1219

\bibitem[{{Mennickent} {et~al.}(2012{\natexlab{a}}){Mennickent},
  {Djura{\v{s}}evi{\'c}}, {Ko{\l}aczkowski}, \& {Michalska}}]{Mennickent2012a}
{Mennickent}, R.~E., {Djura{\v{s}}evi{\'c}}, G., {Ko{\l}aczkowski}, Z., \&
  {Michalska}, G. 2012{\natexlab{a}}, \mnras, 421, 862

\bibitem[{{Mennickent} {et~al.}(2025){Mennickent}, {Djura{\v{s}}evi{\'c}},
  {Rosales}, {Garc{\'e}s}, {Petrovi{\'c}}, {Schleicher}, {Jurkovic},
  {Soszy{\'n}ski}, \&
  {Fern{\'a}ndez-Trincado}}]{Mennickent+2025A&A...693A.217M}
{Mennickent}, R.~E., {Djura{\v{s}}evi{\'c}}, G., {Rosales}, J.~A., {et~al.}
  2025, \aap, 693, A217

\bibitem[{{Mennickent} {et~al.}(2020{\natexlab{a}}){Mennickent}, {Garc{\'e}s},
  {Djura{\v{s}}evi{\'c}}, {Iwanek}, {Schleicher}, {Poleski}, \&
  {Soszy{\'n}ski}}]{2020A&A...641A..91M}
{Mennickent}, R.~E., {Garc{\'e}s}, J., {Djura{\v{s}}evi{\'c}}, G., {et~al.}
  2020{\natexlab{a}}, \aap, 641, A91

\bibitem[{{Mennickent} {et~al.}(2020{\natexlab{b}}){Mennickent}, {Garc{\'e}s},
  {Djura{\v{s}}evi{\'c}}, {Iwanek}, {Schleicher}, {Poleski}, \&
  {Soszy{\'n}ski}}]{Mennickent+2020A&A...641A..91M}
{Mennickent}, R.~E., {Garc{\'e}s}, J., {Djura{\v{s}}evi{\'c}}, G., {et~al.}
  2020{\natexlab{b}}, \aap, 641, A91

\bibitem[{{Mennickent} {et~al.}(2012{\natexlab{b}}){Mennickent},
  {Ko{\l}aczkowski}, {Djurasevic}, {Niemczura}, {Diaz}, {Cur{\'e}}, {Araya}, \&
  {Peters}}]{Mennickent2012b}
{Mennickent}, R.~E., {Ko{\l}aczkowski}, Z., {Djurasevic}, G., {et~al.}
  2012{\natexlab{b}}, \mnras, 427, 607

\bibitem[{{Mennickent} {et~al.}(2008){Mennickent}, {Ko{\l}aczkowski},
  {Michalska}, {Pietrzy{\'n}ski}, {Gallardo}, {Cidale}, {Granada}, \&
  {Gieren}}]{Mennickent2008}
{Mennickent}, R.~E., {Ko{\l}aczkowski}, Z., {Michalska}, G., {et~al.} 2008,
  \mnras, 389, 1605

\bibitem[{{Mennickent} {et~al.}(2016){Mennickent}, {Otero}, \&
  {Ko{\l}aczkowski}}]{Mennickent+2016MNRAS.455.1728M}
{Mennickent}, R.~E., {Otero}, S., \& {Ko{\l}aczkowski}, Z. 2016, \mnras, 455,
  1728

\bibitem[{{Mennickent} {et~al.}(2003){Mennickent}, {Pietrzy{\'n}ski}, {Diaz},
  \& {Gieren}}]{Mennickent+2003}
{Mennickent}, R.~E., {Pietrzy{\'n}ski}, G., {Diaz}, M., \& {Gieren}, W. 2003,
  \aap, 399, L47

\bibitem[{{Michalska} {et~al.}(2010){Michalska}, {Mennickent},
  {Ko{\l}aczkowski}, \& {Djura{\v{s}}evi{\'c}}}]{Michalska2009}
{Michalska}, G., {Mennickent}, R.~E., {Ko{\l}aczkowski}, Z., \&
  {Djura{\v{s}}evi{\'c}}, G. 2010, Astronomical Society of the Pacific
  Conference Series, Vol. 435, {Light Curves of Galactic Eclipsing Double
  Periodic Variables}, ed. A.~{Pr{\v{s}}a} \& M.~{Zejda}, 357

\bibitem[{{Peters}(1994)}]{Peters1994}
{Peters}, G.~J. 1994, in Astronomical Society of the Pacific Conference Series,
  Vol.~56, Interacting Binary Stars, ed. A.~W. {Shafter}, 384

\bibitem[{{Pojmanski}(2002)}]{Pojmanski2002}
{Pojmanski}, G. 2002, \actaa, 52, 397

\bibitem[{{Popper}(1962)}]{Popper1962PASP...74..129P}
{Popper}, D.~M. 1962, \pasp, 74, 129

\bibitem[{{Pringle} {et~al.}(1985){Pringle}, {Wade}, \& {King}}]{Pringle_1985}
{Pringle}, J.~E., {Wade}, R.~A., \& {King}, A. 1985, The Observatory, 105, 241

\bibitem[{{Richards} \&
  {Albright}(1999)}]{Richards&Albright1999ApJS..123..537R}
{Richards}, M.~T. \& {Albright}, G.~E. 1999, \apjs, 123, 537

\bibitem[{{Rosales} \&
  {Mennickent}(2018)}]{Rosales&Mennickent2018IBVS.6248....1R}
{Rosales}, J.~A. \& {Mennickent}, R.~E. 2018, Information Bulletin on Variable
  Stars, 6248, 1

\bibitem[{{Rosales} {et~al.}(2023){Rosales}, {Mennickent},
  {Djura{\v{s}}evi{\'c}}, {Araya}, {Cur{\'e}}, {Schleicher}, \&
  {Petrovi{\'c}}}]{Rosales+2023A&A...670A..94R}
{Rosales}, J.~A., {Mennickent}, R.~E., {Djura{\v{s}}evi{\'c}}, G., {et~al.}
  2023, \aap, 670, A94

\bibitem[{{Rosales} {et~al.}(2024){Rosales}, {Petrovi{\'c}}, {Mennickent},
  {Schleicher}, {Djura{\v{s}}evi{\'c}}, \&
  {Leigh}}]{Rosales+2024A&A...689A.154R}
{Rosales}, J.~A., {Petrovi{\'c}}, J., {Mennickent}, R.~E., {et~al.} 2024, \aap,
  689, A154

\bibitem[{{Sahade} \& {Cesco}(1945)}]{Sahade&Cesco1945}
{Sahade}, J. \& {Cesco}, C.~U. 1945, \apj, 101, 235

\bibitem[{{Sahade} {et~al.}(1997){Sahade}, {Ferrer}, {Garcia}, {Brandi}, \&
  {Barba}}]{Sahade+1997}
{Sahade}, J., {Ferrer}, O., {Garcia}, L.~G., {Brandi}, E., \& {Barba}, R. 1997,
  \pasp, 109, 1237

\bibitem[{{Sahade} \& {Ferrer}(1982)}]{Sahade&Ferrer1982PASP...94..113S}
{Sahade}, J. \& {Ferrer}, O.~E. 1982, \pasp, 94, 113

\bibitem[{{Schleicher} \& {Mennickent}(2017)}]{Schleicher2017}
{Schleicher}, D. R.~G. \& {Mennickent}, R.~E. 2017, \aap, 602, A109

\bibitem[{{Smith} {et~al.}(2010){Smith}, {Henden}, \& {Terrell}}]{Smith2010}
{Smith}, T.~C., {Henden}, A., \& {Terrell}, D. 2010, Society for Astronomical
  Sciences Annual Symposium, 29, 45

\bibitem[{{Spruit} \& {Ritter}(1983)}]{Spruit&Ritter1983A&A...124..267S}
{Spruit}, H.~C. \& {Ritter}, H. 1983, \aap, 124, 267

\bibitem[{{Stellingwerf}(1978)}]{Stellingwerf1978}
{Stellingwerf}, R.~F. 1978, \apj, 224, 953

\bibitem[{{van Leeuwen}(2007)}]{Hipparcos2007}
{van Leeuwen}, F. 2007, \aap, 474, 653

\bibitem[{{VanderPlas}(2018)}]{VanderPlas2018}
{VanderPlas}, J.~T. 2018, \apjs, 236, 16

\bibitem[{{Verbunt}(1993)}]{Verbunt1993ARA&A..31...93V}
{Verbunt}, F. 1993, \araa, 31, 93

\bibitem[{{Vivekananda Rao} \& {Sarma}(1998)}]{Vivekananda&Sarma1998}
{Vivekananda Rao}, P. \& {Sarma}, M.~B.~K. 1998, \aaps, 128, 441

\bibitem[{{Wo{\'z}niak} {et~al.}(2004){Wo{\'z}niak}, {Vestrand}, {Akerlof},
  {Balsano}, {Bloch}, {Casperson}, {Fletcher}, {Gisler}, {Kehoe}, {Kinemuchi},
  {Lee}, {Marshall}, {McGowan}, {McKay}, {Rykoff}, {Smith}, {Szymanski}, \&
  {Wren}}]{NSVS2004}
{Wo{\'z}niak}, P.~R., {Vestrand}, W.~T., {Akerlof}, C.~W., {et~al.} 2004, \aj,
  127, 2436

\bibitem[{{Zechmeister} \& {K{\"u}rster}(2009)}]{Zechmeister2009}
{Zechmeister}, M. \& {K{\"u}rster}, M. 2009, \aap, 496, 577

\end{thebibliography}

\onecolumn
\begin{appendix}
\section{Datasets light curves} \label{sec:appendix_lcs}

\begin{figure}[!h]
    \centering
    \includegraphics[width=0.85\columnwidth]{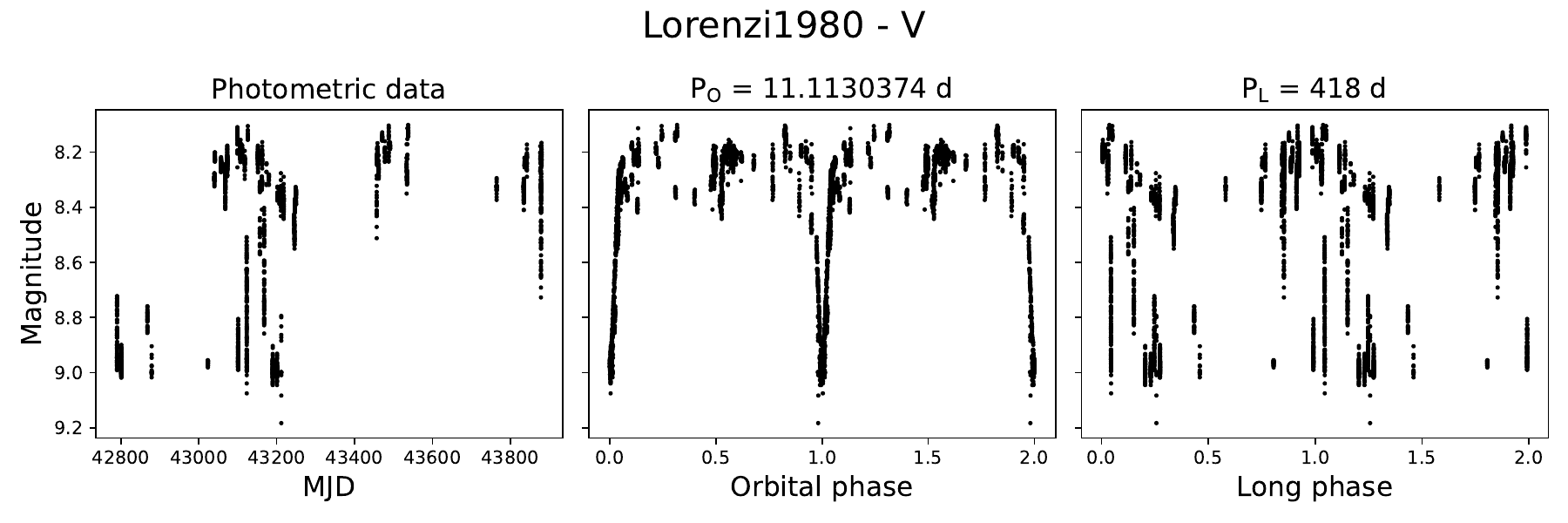}
    \caption{Raw light curve for the Lorenzi1980 dataset, including its phased folded light curves against the orbital and long period reported in D10.}
    \label{fig:appendix_lorenzi1980V}
\end{figure}

\begin{figure}[!h]
    \centering
    \includegraphics[width=0.85\textwidth]{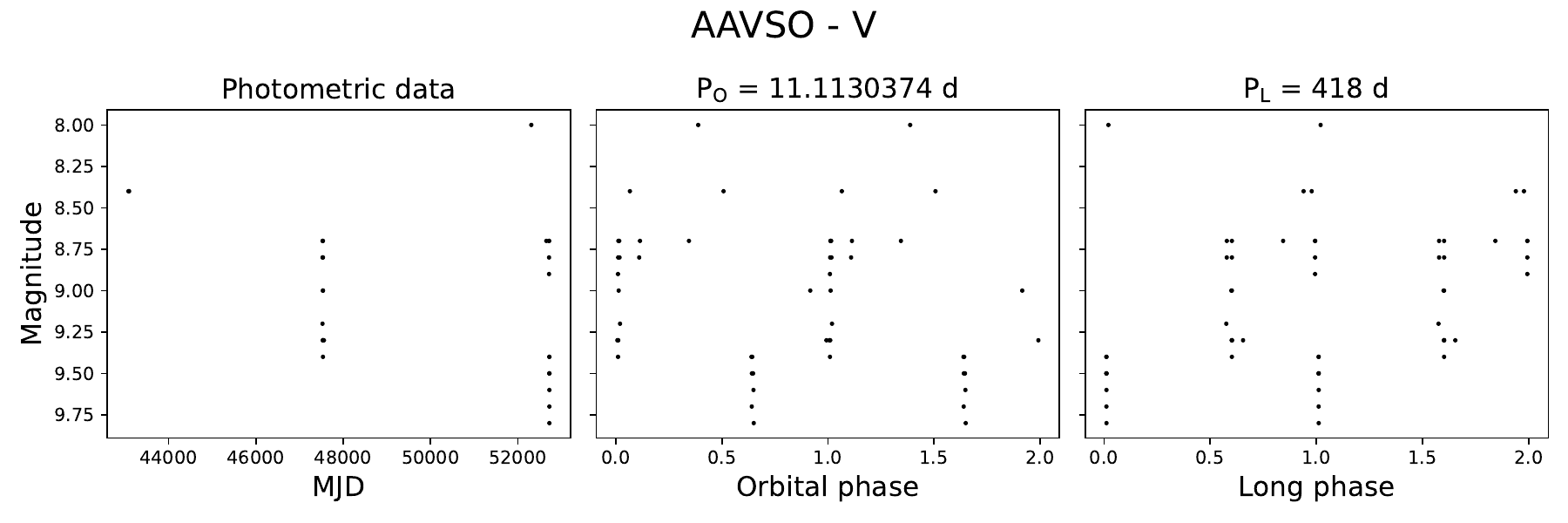}
    \caption{Same as Fig.~\ref{fig:appendix_lorenzi1980V} but for the AAVSO dataset.}
    \label{fig:appendix_aavsoV}
\end{figure}

\begin{figure}[!h]
    \centering
    \includegraphics[width=0.85\textwidth]{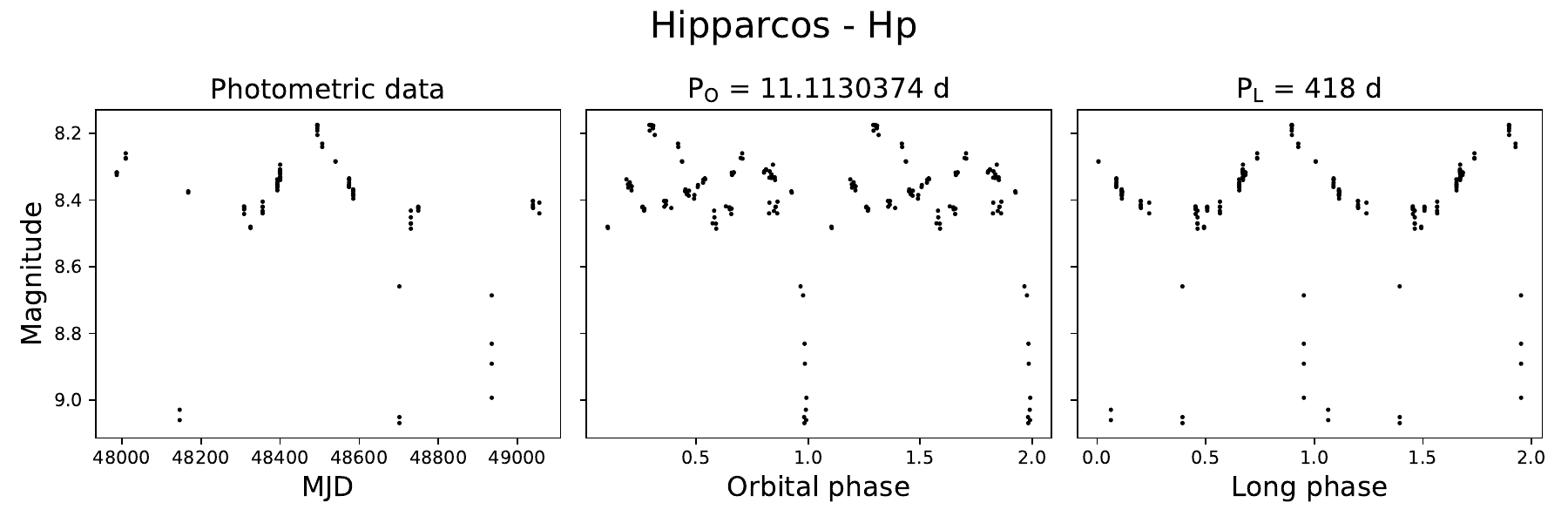}
    \caption{Same as Fig.~\ref{fig:appendix_lorenzi1980V} but for the Hipparcos dataset.}
    \label{fig:appendix_hipparcosHp}
\end{figure}

\begin{figure}[!h]
    \centering
    \includegraphics[width=0.85\textwidth]{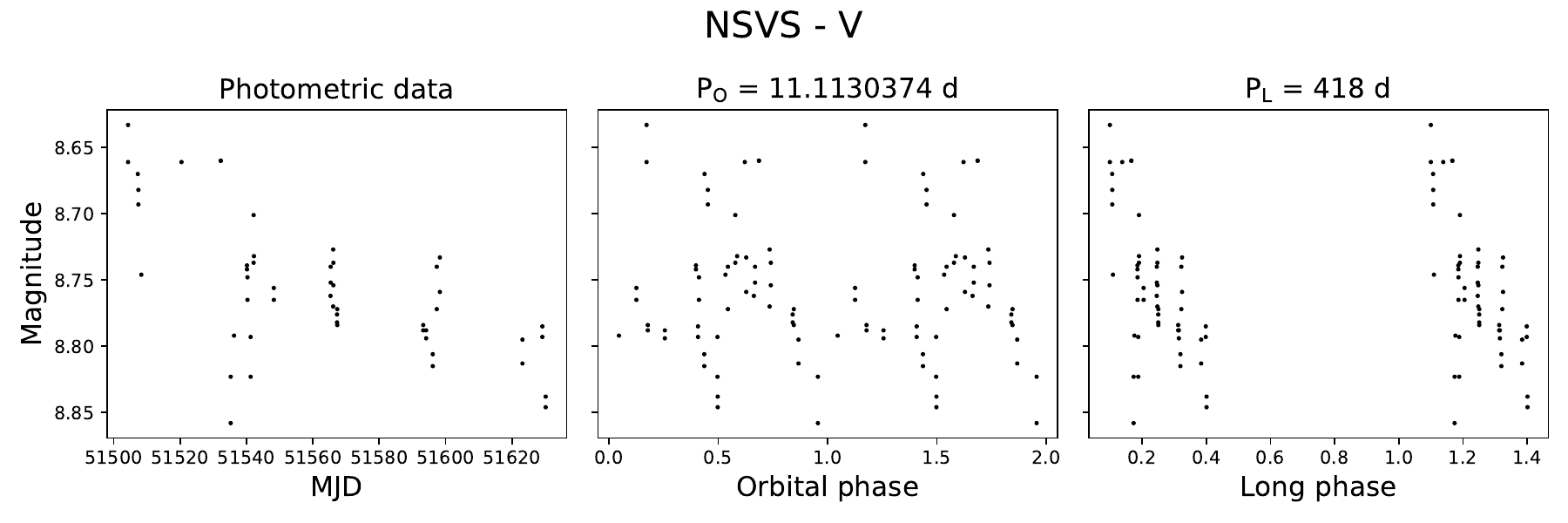}
    \caption{Same as Fig.~\ref{fig:appendix_lorenzi1980V} but for the NSVS dataset.}
    \label{fig:appendix_nsvsV}
\end{figure}

\begin{figure}[!h]
    \centering
    \includegraphics[width=0.85\textwidth]{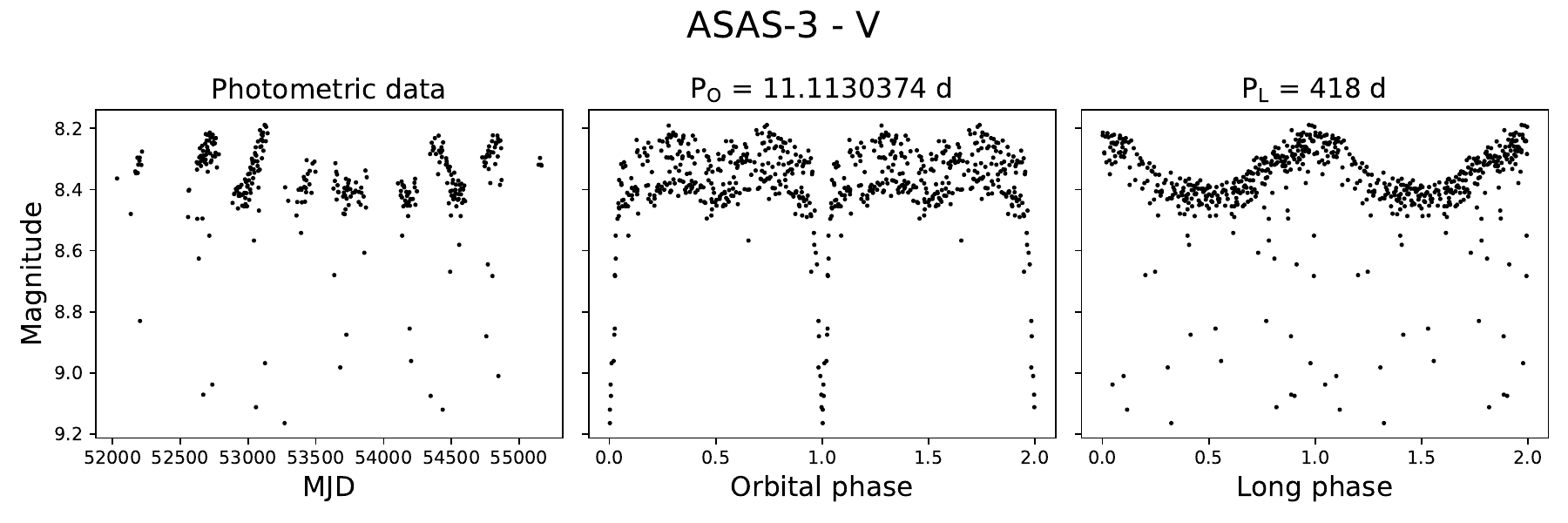}
    \caption{Same as Fig.~\ref{fig:appendix_lorenzi1980V} but for the ASAS-3 dataset.}
    \label{fig:appendix_asas3V}
\end{figure}

\begin{figure}[!h]
    \centering
    \includegraphics[width=0.85\textwidth]{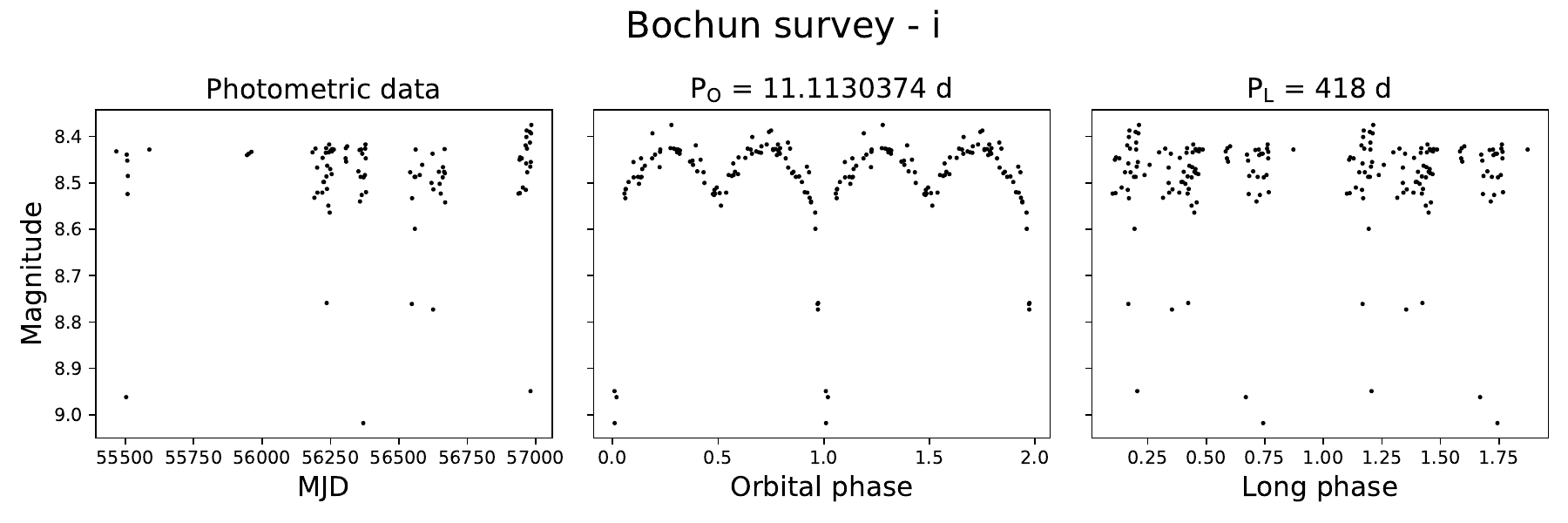}
    \caption{Same as Fig.~\ref{fig:appendix_lorenzi1980V} but for the Bochun dataset.}
    \label{fig:appendix_bochumi}
\end{figure}

\begin{figure}[!h]
    \centering
    \includegraphics[width=0.85\textwidth]{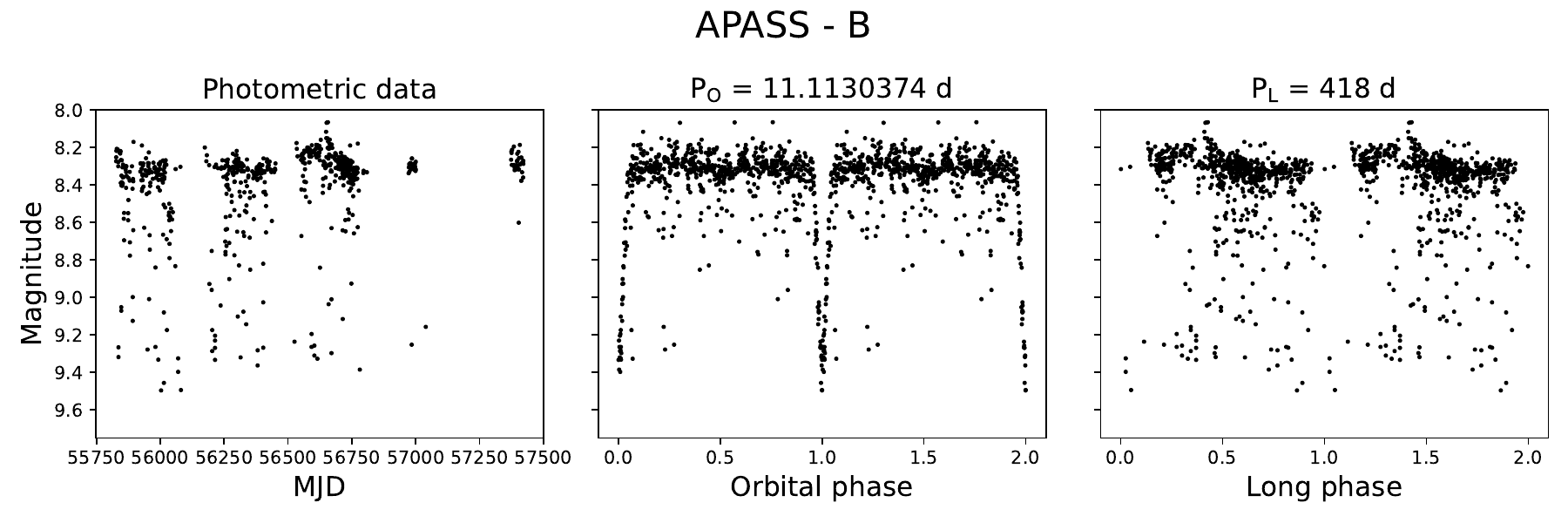}
    \caption{Same as Fig.~\ref{fig:appendix_lorenzi1980V} but for the APASS-B dataset.}
    \label{fig:appendix_apassB}
\end{figure}

\begin{figure}[!h]
    \centering
    \includegraphics[width=0.85\textwidth]{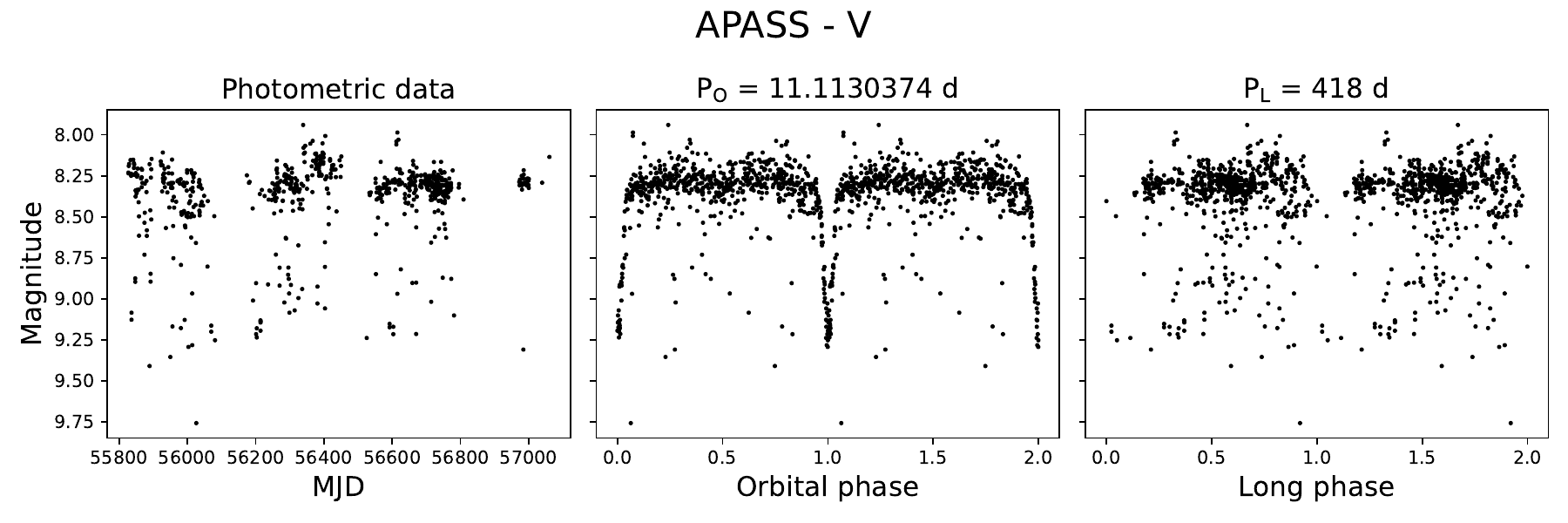}
    \caption{Same as Fig.~\ref{fig:appendix_lorenzi1980V} but for the APASS-V dataset.}
    \label{fig:appendix_apassV}
\end{figure}

\begin{figure}[!h]
    \centering
    \includegraphics[width=0.85\textwidth]{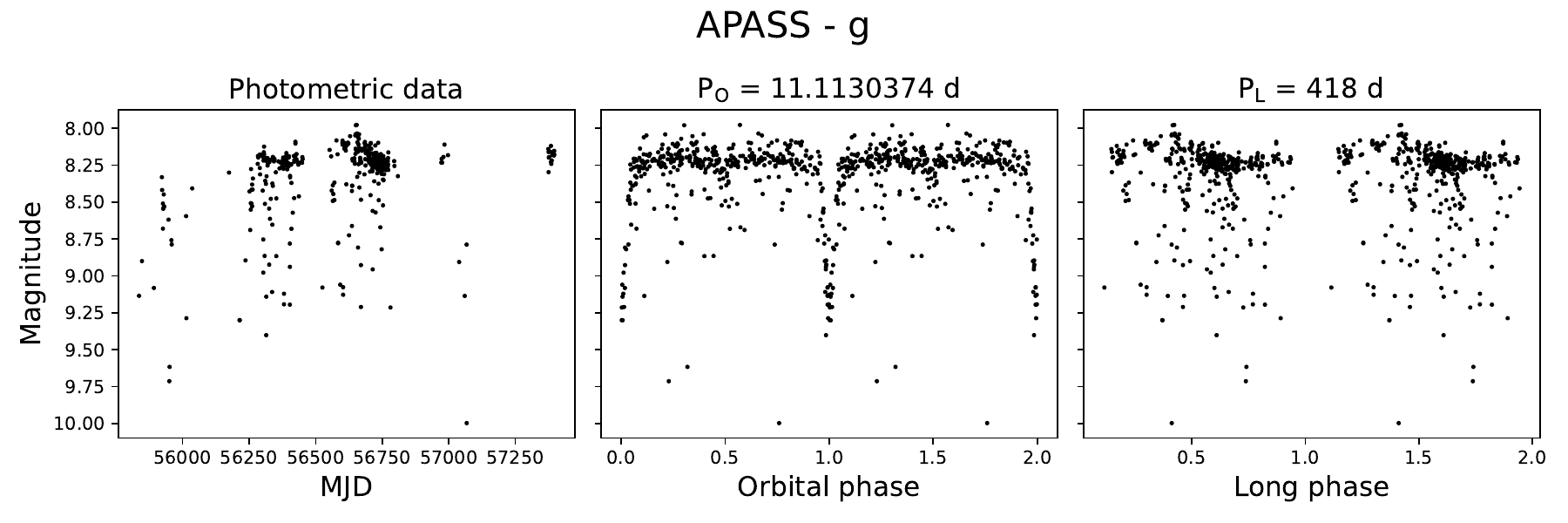}
    \caption{Same as Fig.~\ref{fig:appendix_lorenzi1980V} but for the APASS-g dataset.}
    \label{fig:appendix_apassg}
\end{figure}

\begin{figure}[!h]
    \centering
    \includegraphics[width=0.85\textwidth]{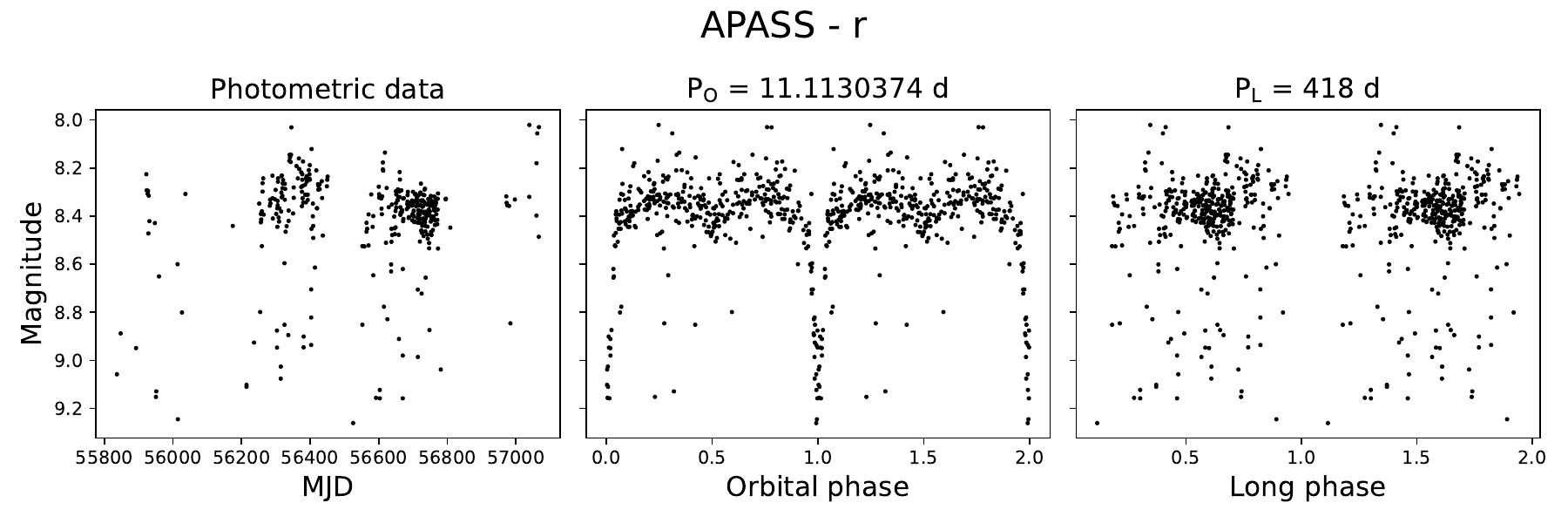}
    \caption{Same as Fig.~\ref{fig:appendix_lorenzi1980V} but for the APASS-r dataset.}
    \label{fig:appendix_apassr}
\end{figure}

\begin{figure}[!h]
    \centering
    \includegraphics[width=0.85\textwidth]{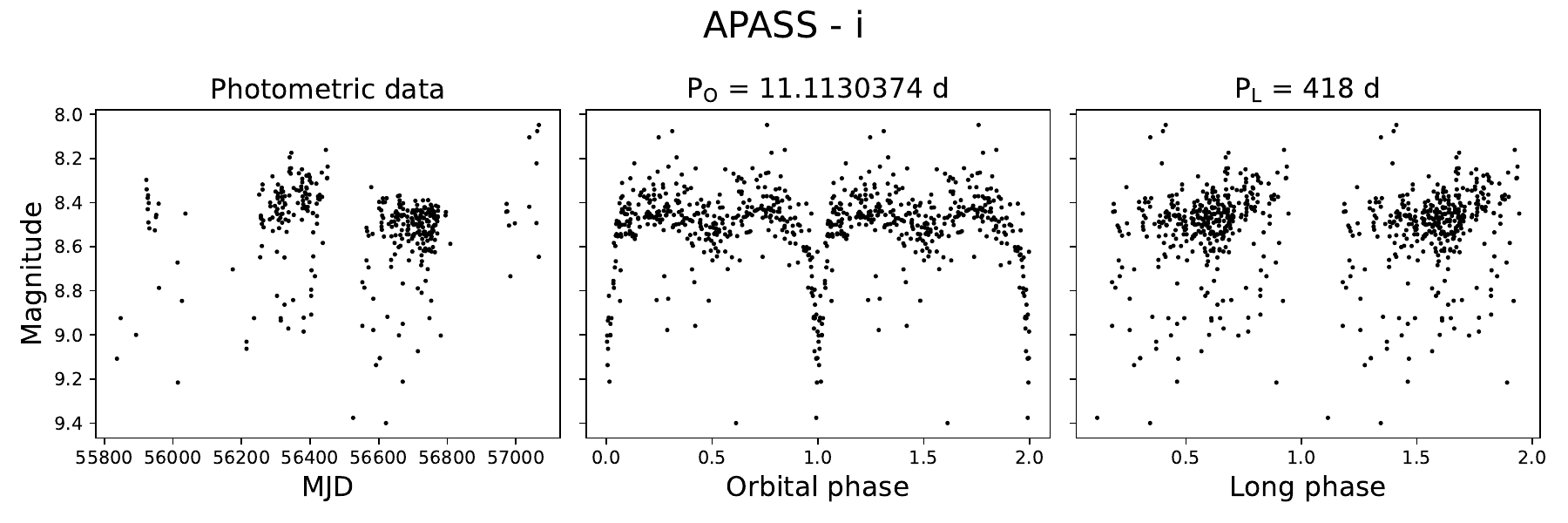}
    \caption{Same as Fig.~\ref{fig:appendix_lorenzi1980V} but for the APASS-i dataset.}
    \label{fig:appendix_apassi}
\end{figure}

\begin{figure}[!h]
    \centering
    \includegraphics[width=0.85\textwidth]{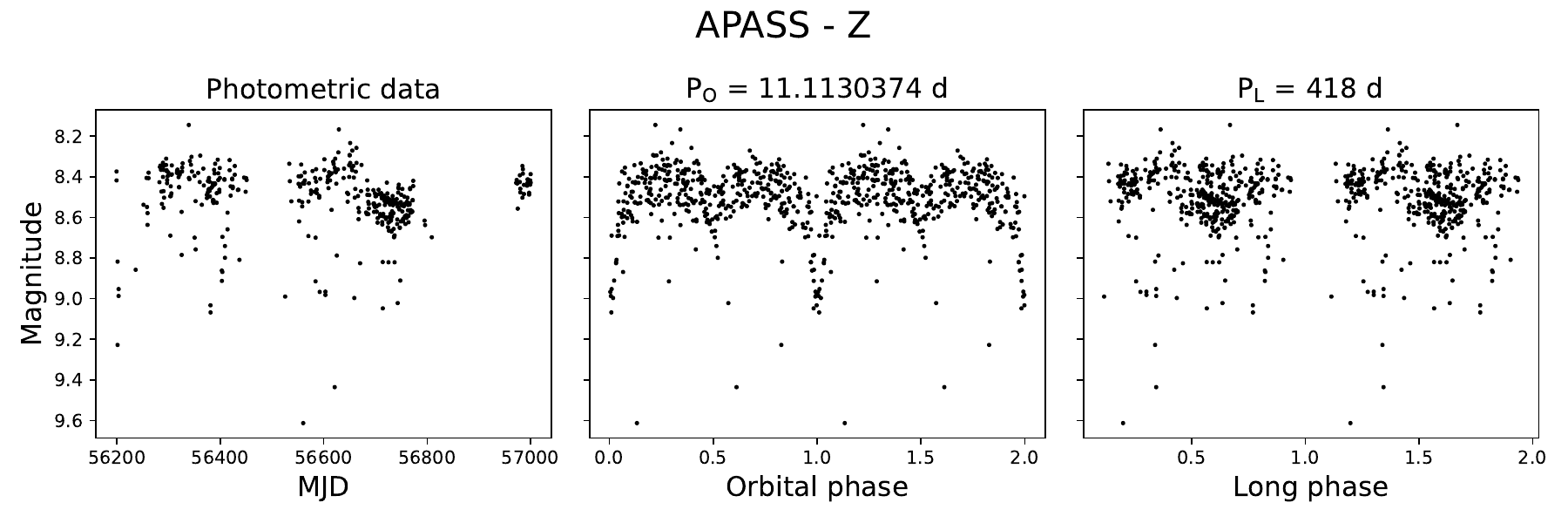}
    \caption{Same as Fig.~\ref{fig:appendix_lorenzi1980V} but for the APASS-Z dataset.}
    \label{fig:appendix_apassZ}
\end{figure}

\begin{figure}[!h]
    \centering
    \includegraphics[width=0.85\textwidth]{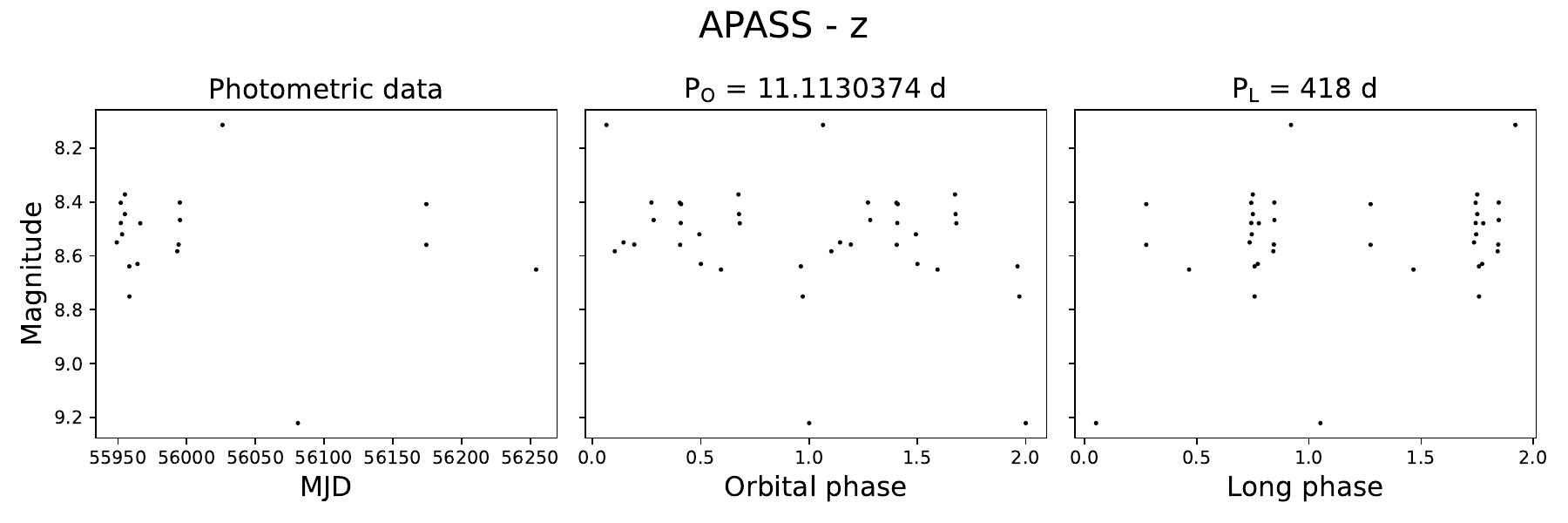}
    \caption{Same as Fig.~\ref{fig:appendix_lorenzi1980V} but for the APASS-z dataset.}
    \label{fig:appendix_apassz}
\end{figure}

\begin{figure}[!h]
    \centering
    \includegraphics[width=0.85\textwidth]{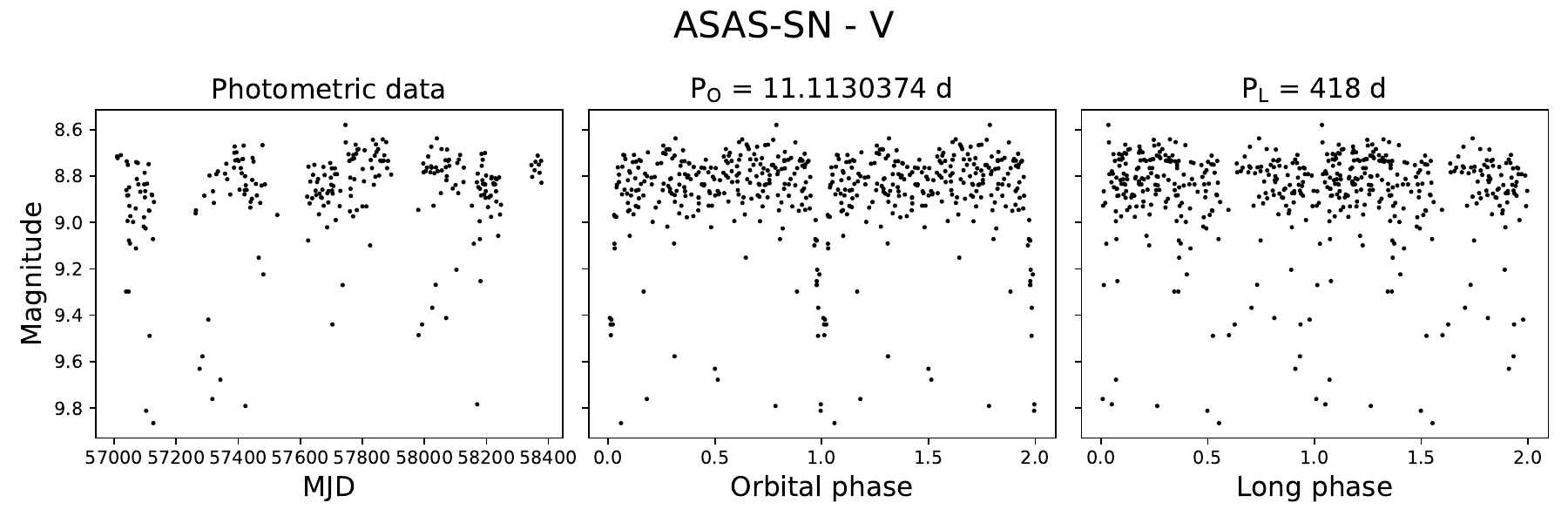}
    \caption{Same as Fig.~\ref{fig:appendix_lorenzi1980V} but for the ASAS-SN dataset.}
    \label{fig:appendix_asassnV}
\end{figure}

\begin{figure}[!h]
    \centering
    \includegraphics[width=0.85\textwidth]{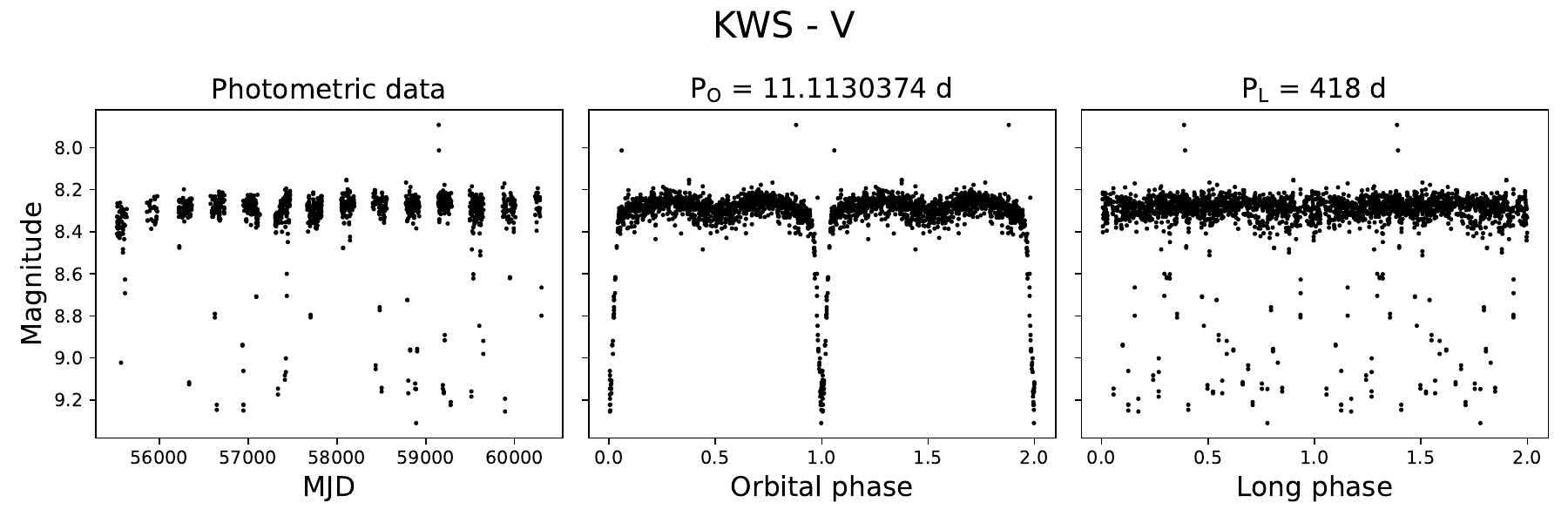}
    \caption{Same as Fig.~\ref{fig:appendix_lorenzi1980V} but for the KWS-V dataset.}
    \label{fig:appendix_kwsV}
\end{figure}

\begin{figure}[!h]
    \centering
    \includegraphics[width=0.85\textwidth]{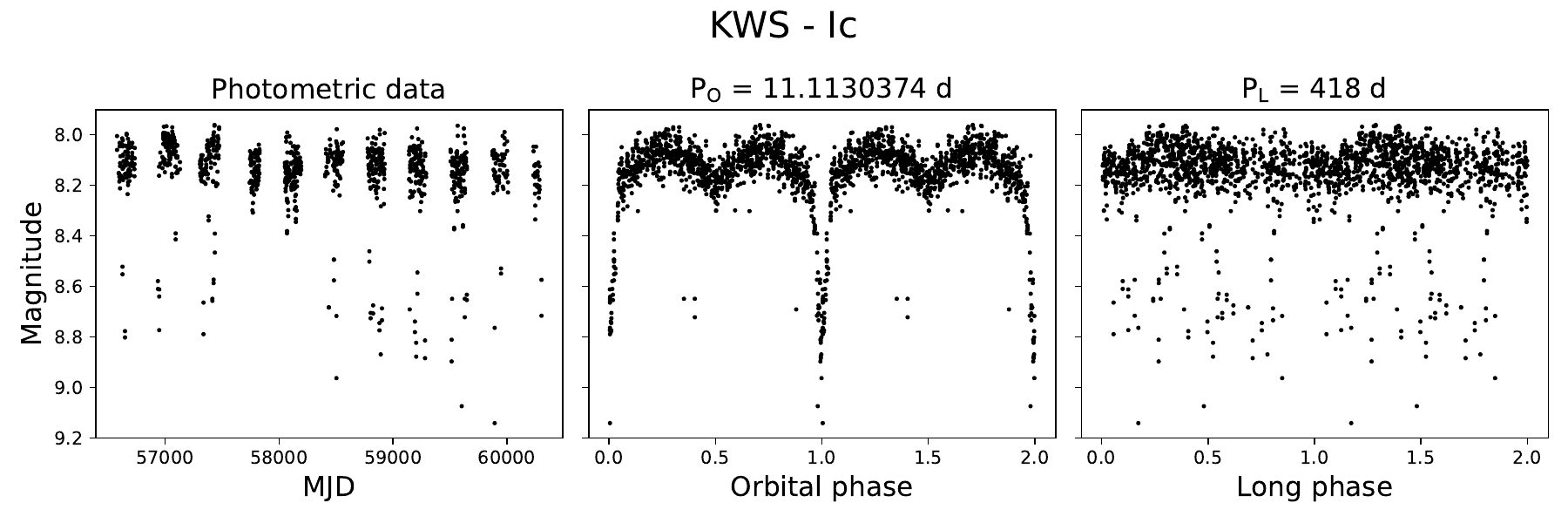}
    \caption{Same as Fig.~\ref{fig:appendix_lorenzi1980V} but for the KWS-I dataset.}
    \label{fig:appendix_kwsIc}
\end{figure}

\end{appendix}
\end{document}